\documentclass[aps,pra,notitlepage,reprint,floatfix,amsmath,amssymb]{revtex4-1}

\usepackage{siunitx}
\usepackage{xcolor}
\usepackage{graphicx}
\graphicspath{{Figures/}}         % Path to search for figures
\usepackage[caption=false]{subfig} % Subfig does not mess with RevTex
\usepackage{hyperref}
\usepackage{bbm}
\usepackage{mathrsfs}

\DeclareMathOperator{\ind}{\mathbbm{1}}

\renewcommand{\vec}[1]{\mathbf{{#1}}}
\newcommand{\pvec}[1]{\mathbf{{#1}}_\parallel }
\newcommand{\vecUnit}[1]{\hat{\mathbf{#1}}}

\newcommand{\dtwox}{{\mathrm{d}^2 x_\parallel}}

\DeclareMathOperator{\ellb}{\boldsymbol{\ell}}

\newcommand{\Vie}[3]{\mathop{\mathbf{#1}_{#2}^{#3}}}
\newcommand{\Cie}[3]{\mathop{\mathcal{#1}_{#2}^{#3}}}

\DeclareMathOperator{\eqnume}{\, \buildrel \mathrm{num} \over = \,}
\DeclareMathOperator{\eqexp}{\, \buildrel \mathrm{exp} \over = \,}

\begin{document}

%----------------------------------------------------------------------
%  TITLE and AUTHORS
%----------------------------------------------------------------------

\title{Haze and light diffraction of dielectric line-gratings}
\author{Jean-Philippe Banon$^{1,2,3}$}
\author{Colette Turbil$^{1}$}
\author{Iryna Gozhyk$^{1}$}
\author{Ingve Simonsen$^{1,4}$}
\affiliation{$^1$Surface du Verre et Interfaces, UMR 125 CNRS/Saint-Gobain, F-93303 Aubervilliers, France}
\affiliation{$^2$Institut Langevin, ESPCI Paris, CNRS, PSL University, 1 rue Jussieu, F-75005 Paris, France}
\affiliation{$^3$Laboratoire de Physique de la Matière Condensée, CNRS, Ecole Polytechnique, Institut Polytechnique de Paris, 91120 Palaiseau, France}
\affiliation{$^4$Department of Physics, NTNU -- Norwegian University of Science and Technology, NO-7491 Trondheim, Norway}

\date{\today}

%----------------------------------------------------------------------
%  ABSTRACT
%----------------------------------------------------------------------
\begin{abstract}
The optical characterization of dielectric periodic line-gratings for which the period of the grating is significantly larger than the probing wavelength is investigated experimentally, numerically, and analytically. The experimental diffraction efficiencies are analyzed in transmission with the help of a closed-form approximate solution within a single scattering picture. In particular, the angular width of significant scattering and the haze are analyzed for samples of increasing grating profile amplitude. 
\end{abstract}

\maketitle % PRA

%-------------------------------------------------------------------------
\section{Introduction}
%-------------------------------------------------------------------------
Thanks to recent advancements in micro- and nano-fabrication techniques, it is now possible to produce macroscopic objects with surface and volume features down to a scale of hundreds of nano-meters~\cite{Su2018}. Significant efforts of the scientific community have been dedicated to theoretical and experimental studies of the periodic patterns for they are the simplest model cases to analyze and to manufacture. For instance, periodic patterns with sub-micrometer and micro-meter scale lattice constants were extensively studied during the past decade for numerous photonics applications including light extraction, photovoltaics etc. In contrast, patterns with larger lattice constants $a$ of the order of $10^{1}$--$10^{2}$ micrometers were almost omitted, most probably due to the complexity of the analysis they require for application in the visible spectral range. Indeed, patterns with lattice constant to wavelength ratio in the range  $20\leq a/\lambda\leq 2000$ can sustain up to thousands of diffraction orders. The numerical simulations of their optical properties require substantial discretization efforts and are computationally very demanding. The huge number of diffraction orders is often considered as an indicator of low diffraction efficiency for the majority of diffraction orders higher than $5$--$10$, even though this latter assumption remains to be proven for an arbitrary shape of the surface profile. 
Meanwhile, it has been shown that in patterns with periods larger than \SI{9}{\micro\meter} illuminated with visible light the diffraction $m\geq1$ appears within the solid angle of \ang{2.5} around the specularly diffracted ones. As a result in such structures, the interpretation of the integral optical properties such as haze and gloss can be misleading~\cite{Turbil2019}.

\smallskip
The rest of this paper is organized as follows. Section~\ref{sec:exp} presents the set of samples examined in this work and the instruments used to perform the topography analysis and the angle-resolved transmission measurements. Section~\ref{sec:theory} presents the theoretical models applied to analyze the experimental data. Section~\ref{sec:results} is devoted to the experimental and theoretical angle-resolved optical properties of the samples and their comparison.  Finally, the conclusions that can be drawn from the present work are presented in Sec.~\ref{sec:conclusion}.

%-------------------------------------------------------------------------
\section{Sample fabrication and scattering experiment}\label{sec:exp}
%-------------------------------------------------------------------------
\subsection{Sample manufacturing}

Line gratings examined in this work were produced with nanoimprint lithography (NIL,\cite{Chou}) combined with post-lithography annealing of the samples. 
NIL is a fabrication technique that allows the elaboration of patterned dielectric surfaces with sub-micrometer to sub-millimiter lateral dimensions of the pattern elements. An initial master surface with a surface pattern of interest is chosen to be reproduced. As a first step of the NIL process, a liquid silicone is cast on the master surface. The silicone is then thermally cured to turn to its final solid-elastic state and can therefore be demolded from the master. This soft mold is then textured with the negative pattern of the master surface. The soft mold is then pressed upon a liquid film of polymer or sol-gel deposited on a rigid or a flexible substrate. The film thickness and allied pressure are chosen such that the film completely fills the mold while assuring the adhesion with the substrate. The coating is then cross-linked by temperature rise before the release of the mold and a permanent surface morphology, which is identical to the master surface, is thus obtained. The NIL process is easily scalable, and patterned surfaces of tens of centimeters can be produced with high reproducibility~\cite{Dubov,Brudieu2017}. \\

By choosing a thermoplastic polymer material with adapted chemical properties, one can slightly modify the initial pattern. Indeed, Teisseire~\textit{et al.} realized such micrometric patterns thanks to NIL with poly(methyl methacrylate) (PMMA), and were able to slightly modify the structure shape of the pattern upon NIL, by rising its temperature up to the glass transition temperature of the PMMA~\cite{Teisseire2011}. This way, the PMMA material goes back to viscous state, and under the effect of the surface tension forces, the profile is slowly smoothed without impacting the pattern period. This thermal relaxation is stopped by a fast removal of the heating source, leading to the solidification of the PMMA film on its current state. \\

\begin{figure}[t]
\centering
\includegraphics[width = 8cm]{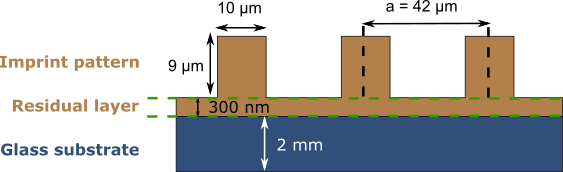}
\caption{Schematic representation of the initial periodic samples: a \SI{2}{mm} thick glass substrate with a micrometric line pattern on its top. Each line is rectangular has height and width of about \SI{10}{\micro\meter}. The lattice constant is around \SI{42}{\micro\meter}. A residual layer of \SI{300}{nm} between the glass substrate and the micrometric pattern is necessary to provide adhesion between the substrate and the lines.}
\label{fig:schema_sample}
\end{figure}

Following this combined NIL-thermal relaxation procedure, six samples with identical micrometric patterns were fabricated on a $5 \times 5$~cm$^2$ large and $2$~mm thick glass substrate, using a solution of PMMA. The solution is obtained by mixing a PMMA powder (25$\%$ by weight) and a 4-Methyl-2-pentanone solution. The geometry of the sample is presented on Fig.~\ref{fig:schema_sample}. The micrometric pattern consists of a line grating of period of $40~\mu$m. The cross-section of each line is rectangular, with height and width of \SI{10}{\micro\meter}. A residual layer is required between the glass substrate and the micrometric pattern in order to provide adhesion between the glass and the lines. In order to avoid strong light guiding effects, the residual layer is manufactured to be \SI{300}{nm} thick.

%-------------------------------------------------------------------------
% --- Figure
%-------------------------------------------------------------------------
%-------------------------------------------------------------------------

\begin{figure*}[t]
\centering
\includegraphics[width = 15cm]{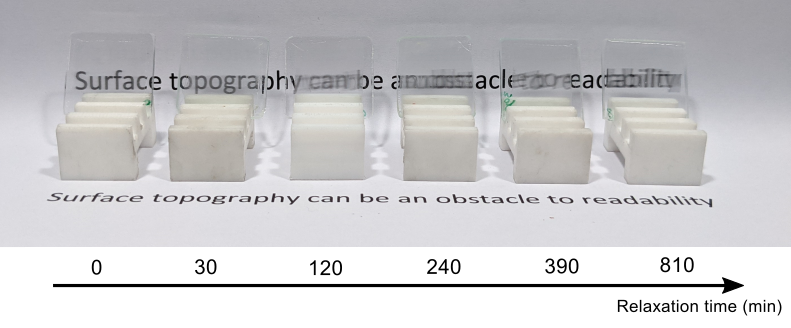}
\caption{Optical photographs of the image blurring due to the presence of examined samples between the observer and the image.}
\label{fig:visual}
\end{figure*}

Six samples are put in the oven at $130^{\circ}$C and are progressively removed from it after \SI{30}{min}, 120 min, 240 min, 390 min and 810 min of thermal treatment. All samples are then cooled down to room temperature. Figure ~\ref{fig:visual} illustrates how the thermal relaxation impacts the visual aspect in transmission of the studied samples. For annealing durations higher than 120 minutes, image blurring appears, and the text behind the samples are not readable anymore.
%-------------------------------------------------------------------------
% --- Figure
%-------------------------------------------------------------------------
%-------------------------------------------------------------------------

\subsection{Morphological characterization} 

\begin{figure}[t]
\begin{center}
%\begin{subfigure}[b]{0.48\textwidth}
\includegraphics[width=.45\textwidth,trim = 0cm 0cm 0cm 0cm]{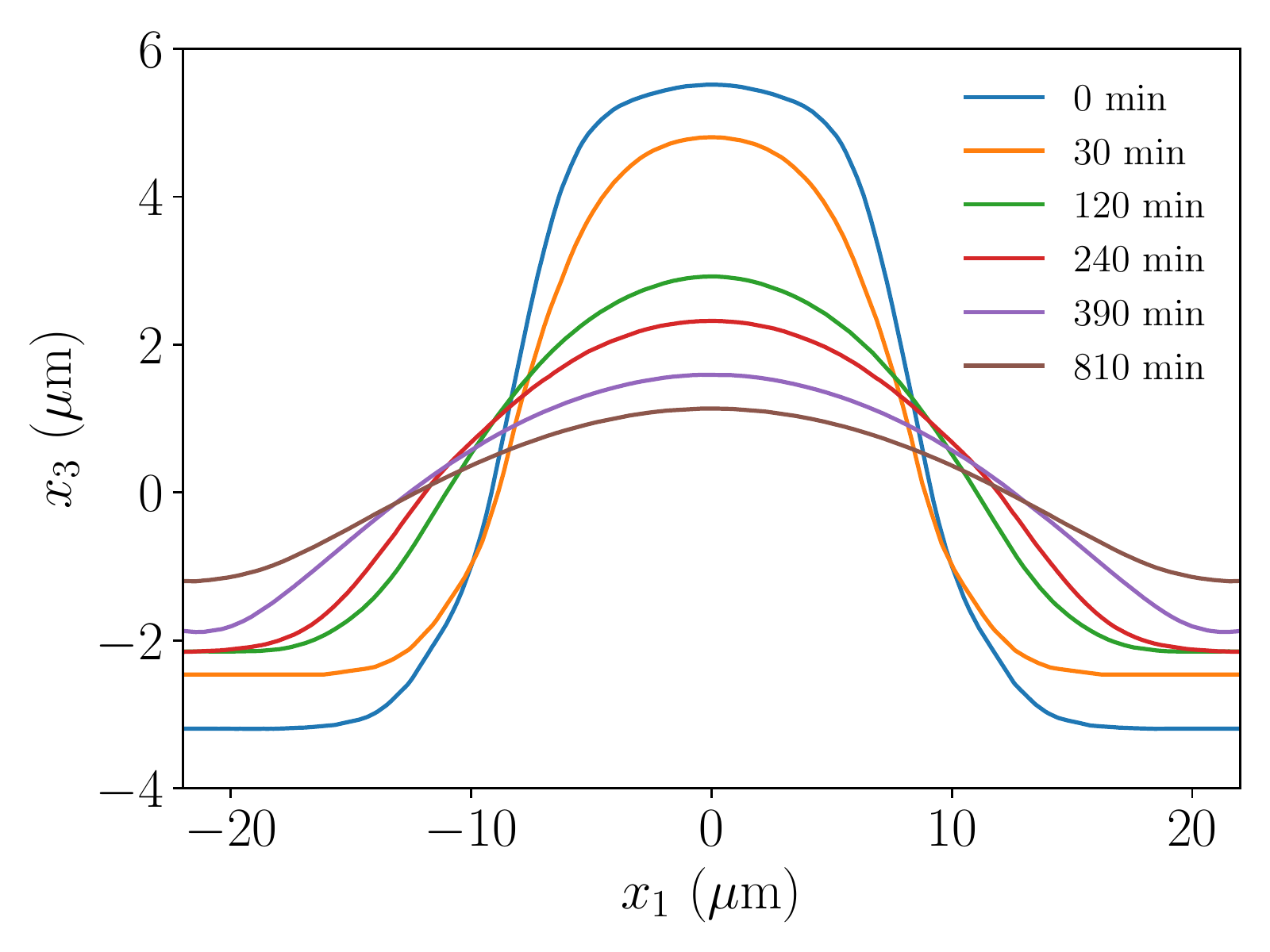}
\caption{Surface profiles of the samples obtained after different relaxation times ranging from 0 to 810 minutes.}
\label{fig:profile}
\end{center}
\end{figure}

The profiles of the six surfaces are measured with a contact profilometer (Dektak, Bruker). The 1D surface profiles are measured along the direction that is assumed to be orthogonal to the line pattern (with misalignment of less than 5 degrees) and are presented on Fig.~\ref{fig:profile}. The length of measured 1D profiles were \SI{600}{\micro\meter} and measurement step was \SI{0.03}{\micro\meter}, but for the sake of clarity a single period for the average profile is shown on Fig.~\ref{fig:profile}.

%-------------------------------------------------------------------------
% --- Figure
%-------------------------------------------------------------------------
%-------------------------------------------------------------------------

\subsection{Optical measurements} 
The integral optical properties such as haze and clarity were measured with hazemeter "Haze-gard plus" commercialized by BYK Gardner, while the angle-resolved light scattering experiments were performed with the goniospectrophotometer device OMS4 commercialized by the OPTIS company. 
OMS4 consists of a sample holder, light sources and a photomultiplicator detector, which with the help of two precise motors, can scan the whole space around the sample and thus measure the angular distribution of the scattered light (see Ref.~\onlinecite{Turbil2019} for details).
\\

\begin{figure*}[t]
\begin{center}
\includegraphics[width=.9\textwidth,trim = 0cm 0cm 0cm 0cm]{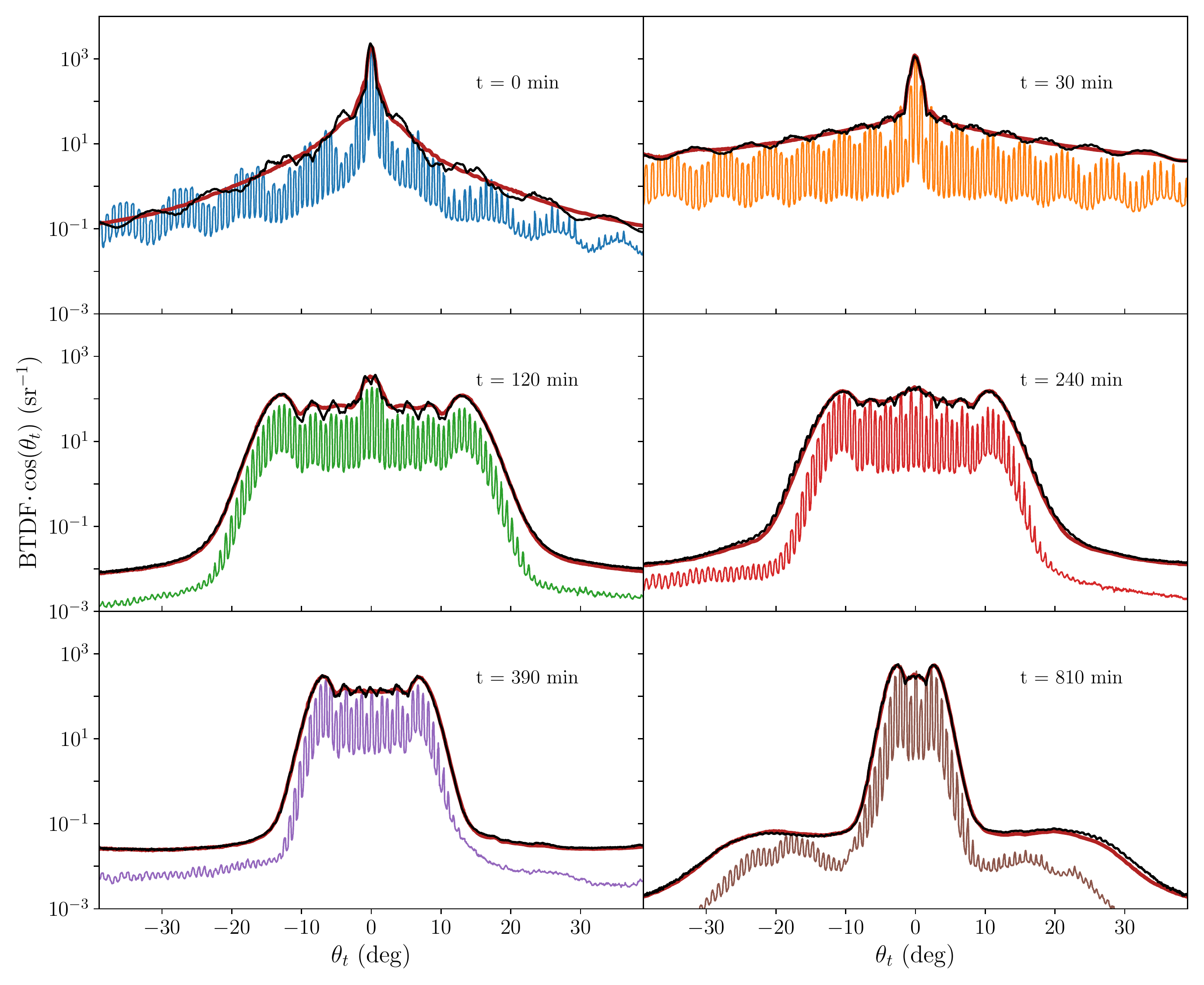}
\caption{Cosine-corrected BTDF as a function of the angle of transmission obtained at normal incidence. The measurements were performed with laser beam of wavelength $\lambda = \SI{520}{nm}$ (oscillatory solid lines of different colors), a broad-band source (smooth solid red lines), or a broad-band source filtered at \SI{535}{nm} and using a filter width on \SI{10}{nm}~(solid black lines). Both illumination sources were p-polarized.}
%\caption{Cosine-corrected BTDF as a function of the angle of transmission obtained at normal incidence. Measurements preformed with laser beam (solid line),  broad-band source (solid brown line) or broad-band source filtered at \SI{535}{nm} with filter width on \SI{10}{nm}~(solid black line). The wavelength of the laser beam was $\lambda = \SI{520}{nm}$. Both illumination sources were p-polarized.}
\label{fig:btdf}
\end{center}
\end{figure*}

Samples were illuminated at normal incidence by a p-polarized green laser (wavelength \SI{520}{nm}) or by a broad-band source. Here by p-polarized light, we mean that the magnetic field is parallel to the axis of invariance of the grating. Only the light transmitted through the sample was measured. In measurements with laser source, a diaphragm of \SI{1.05}{mm} diameter was put in front of the photomultiplicator, to control the solid angle of the detection and ensure an angular resolution of \ang{0.16}, while in measurements with broad-band source, a diaphragm of 4 mm diameter was used, resulting in angular resolution of \ang{0.5}. As such one-dimensional patterned surfaces only scatter light in one plane for non-conical incidence, measurements are achieved from $-\ang{40}$ to \ang{40} around the normal direction, with an angular detection step of \ang{0.05}. Measurements are presented on Fig.~\ref{fig:btdf} for six samples exhibiting different bidirectional transmission distribution functions~(BTDF).

%-------------------------------------------------------------------------
% --- Figure
%-------------------------------------------------------------------------
%-------------------------------------------------------------------------

%-------------------------------------------------------------------------
% --- Figure
%-------------------------------------------------------------------------
\begin{figure}[t]
\begin{center}
\includegraphics[width=.45\textwidth,trim = 0cm 0cm 0cm 0cm]{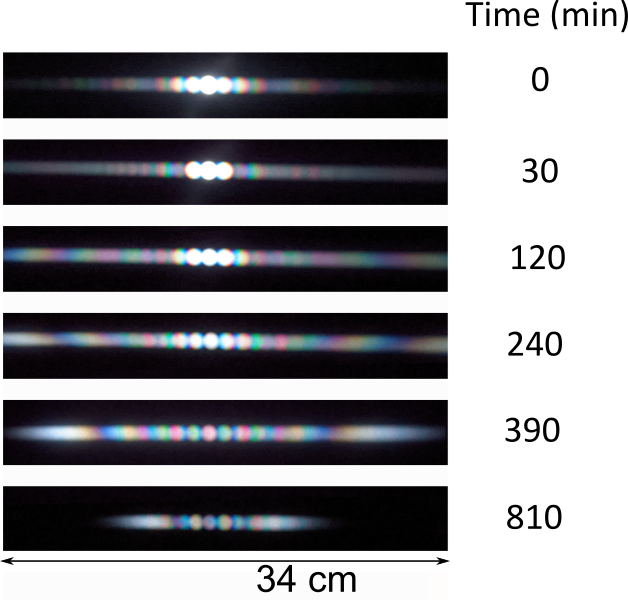}
\caption{Optical photography taken during the samples alignment within BTDF measurements. Light from broad-band source scattered through the sample in transmission is visualized with the help of black screen placed at about 2 m distance from the sample.IG: to add the size. }
\label{fig:photo}
\end{center}
\end{figure}
%-------------------------------------------------------------------------

%-------------------------------------------------------------------------
\section{Theory}\label{sec:theory}
%-------------------------------------------------------------------------
\subsection{Reduced Rayleigh equations} 
The theoretical framework used here for analyzing the diffraction efficiencies is that of the reduced Rayleigh equations for periodic systems~\cite{kretschmann2002,banon:thesis}.
% The gratings studied experimentally are one-dimensional (invariance of the profile along the $x_2$ direction) and illuminated under normal incidence, which is a particular case of non-conical incidence (the direction of the incident wave vector belongs to the symmetry plane of the grating, $(x_1,x_3)$-plane). Under such conditions, the wave vectors of the diffracted modes belong to the plane of symmetry. The $x_1$-component of such a wave vector will be denoted $p_{\ell}$ with $\ell \in \mathbb{N}$ indexing a diffracted order and is given by the grating formula
%%
%\begin{equation}
%p_{\ell} = p_{0} + \frac{2 \pi \ell }{a} \: ,
%\end{equation}
%%
%where $p_0$ is the $x_1$-component of the incident wave vector.
For a periodic scattering system, and for a given incident plane wave, a reduced Rayleigh equation is an infinite countable linear system of equations whose unknown are the reflection or transmission amplitudes appearing in the plane wave expansion of the scattered electric field. In other words, for a monochromatic plane wave incident on the sample, of the form
\begin{equation}
\Vie{E}{0}{} (\Vie{r}{}{}) = \sum_{\nu = p, s} \mathcal{E}_{0,\nu} \, \Vie{\hat{e}}{1,\nu}{-} (\Vie{p}{0}{}) \, \exp \Big[ i \Vie{k}{1}{-}(\Vie{p}{0}{}) \cdot \Vie{r}{}{} \Big] \: ,
\end{equation}
the transmitted scattered far-field may be written as
\begin{align}
\Vie{E}{j}{} (\Vie{r}{}{}) = & \sum_{\ellb \in \mathbb{Z}^2}  \: \sum_{\mu = p,s} \Vie{\hat{e}}{\mu,j}{-} (\Vie{p}{\ellb}{}) \sum_{\nu = p,s} T_{\mu \nu}^{(\ellb)} (\Vie{p}{0}{}) \, \mathcal{E}_{0,\nu} \nonumber\\
&\times \exp \Big[ i \Vie{k}{j}{-}(\Vie{p}{\ellb}{}) \cdot \Vie{r}{}{} \Big] \: .
\end{align}
Here an overall phase factor $\exp (-i \omega t)$ is implicitly assumed and dropped for clarity. The index $j$ refers to the medium of transmission. In our case, we will model the scattering system either as a surface profile separating two semi-infinite homogeneous media of dielectric constants $\varepsilon_1$ and $\varepsilon_2$ or a system made of three media if the planar bottom interface of the substrate is considered (hence $j=2$ for a scattering system made of two media, $j = 3$ for three media). The dispersion relation in each medium reads $|\Vie{k}{}{}| = \sqrt{\epsilon_j} \, \omega / c \equiv k_j$ ($j \in \{1,2,3\}$). The wave vectors $\Vie{p}{\ellb}{}$ and $\Vie{k}{j}{\pm}(\Vie{p}{\ellb}{})$ are defined by
\begin{subequations}
\begin{align}
\Vie{p}{\ellb}{} &= \Vie{p}{0}{} + \Vie{G}{\ellb}{} \\
\alpha_j (\Vie{p}{\ellb}{}) &= \left( \epsilon_j \frac{\omega^2}{c^2} - \Vie{p}{\ellb}{2} \right)^{1/2}, \: \mathrm{Re} (\alpha_j), \: \mathrm{Im} (\alpha_j) \geq 0 \\
\Vie{k}{j}{\pm}(\Vie{p}{\ellb}{}) &= \Vie{p}{\ellb}{} \pm \alpha_j (\Vie{p}{\ellb}{}) \, \Vie{\hat{e}}{3}{} \: ,
\end{align}
\end{subequations}
and the $s$ and $p$ polarization vectors are defined by
\begin{subequations}
\begin{align}
\Vie{\hat{e}}{j,s}{\pm} (\Vie{p}{\ellb}{}) &\equiv \Vie{\hat{e}}{s}{} (\Vie{p}{\ellb}{}) = \frac{\Vie{\hat{e}}{3}{} \times \Vie{k}{j}{\pm}(\Vie{p}{\ellb}{})}{|\Vie{\hat{e}}{3}{} \times \Vie{k}{j}{\pm}(\Vie{p}{\ellb}{}) |}  \: ,\\
\Vie{\hat{e}}{j,p}{\pm} (\Vie{p}{\ellb}{}) & = \frac{\Vie{\hat{e}}{s}{} (\Vie{p}{\ellb}{}) \times \Vie{k}{j}{\pm}(\Vie{p}{\ellb}{})}{|\Vie{\hat{e}}{s}{} (\Vie{p}{\ellb}{}) \times \Vie{k}{j}{\pm}(\Vie{p}{\ellb}{})|} \: .
\end{align}
\end{subequations}
Here $\Vie{p}{0}{} = k_1 \, \sin \theta_0 \, (\cos \phi_0 \, \Vie{\hat{e}}{1}{} + \sin \phi_0 \, \Vie{\hat{e}}{2}{})$ is the projection of the wave vector of the incident plane wave, $\Vie{k}{1}{-}(\Vie{p}{0}{})$, into the $(\Vie{\hat{e}}{1}{}, \Vie{\hat{e}}{2}{})$-plane. The vector $\Vie{G}{\ellb}{} = \ell_1 \, \Vie{b}{1}{} + \ell_2 \, \Vie{b}{2}{}$ is a reciprocal lattice vector (with $\Vie{a}{i}{} \cdot \Vie{b}{j}{} = 2 \pi \delta_{ij}$ and $\Vie{a}{i}{}$ a primitive lattice vector). The vectors $\Vie{k}{j}{\pm}(\Vie{p}{\ellb}{})$, $\Vie{\hat{e}}{j,p}{\pm} (\Vie{p}{\ellb}{})$ and $\Vie{\hat{e}}{s}{} (\Vie{p}{\ellb}{})$ thus represent the wave vector of a plane wave scattered in medium $j$ and propagating or decaying exponentially away from the surface and the associated $p$ and $s$ local polarization basis vectors, respectively.
The complex amplitude $T_{\mu \nu}^{(\ellb)}( \Vie{p}{0}{})$ for $\mu, \nu \in \{p,s\}$ is the transmission amplitude for the transmitted plane wave with lateral wave vector $\Vie{p}{\ellb}{}$ in the polarization state $\mu$ from a unit incident field with lateral wave vector $\Vie{p}{0}{}$  in the polarization state $\nu$.
 It can be shown that for a scattering system made of two media, the transmission amplitudes satisfy the following reduced Rayleigh equation~\cite{Toigo:77,soubret1,hetland:17,Banon:2018,banon:thesis},
\begin{widetext}
\begin{equation}
\sum_{\mathbf{m} \in \mathbb{Z}^2} \Cie{J}{12}{-,-}(\Vie{p}{\ellb}{} , \Vie{p}{\mathbf{m}}{}) \: \Vie{M}{12}{-,-}(\Vie{p}{\ellb}{} , \Vie{p}{\mathbf{m}}{}) \: \mathbf{T}^{(\mathbf{m})} (\Vie{p}{0}{}) = - \frac{2 n_1 n_2 \alpha_1(\Vie{p}{\ellb}{})}{\epsilon_2 - \epsilon_1} \delta_{\ellb, \mathbf{0}} \Vie{I}{2}{} \: ,
\label{RREint}
\end{equation}
for all mode index $\ellb \in \mathbb{Z}^2$. Here $\mathbf{T}^{(\mathbf{m})} (\Vie{p}{0}{}) = (T_{\mu \nu}^{(\mathbf{m})} (\Vie{p}{0}{}))_{\mu \nu \in \{p,s\}}$ and $\Vie{I}{2}{}$ is the 2$\times$2 identity matrix. The matrix $\Vie{M}{jk}{b,a} (\Vie{p}{\ellb}{},\Vie{p}{\mathbf{m}}{})$ is the 2$\times$2 matrix defined by
\begin{equation}
	\Vie{M}{jk}{b,a} (\Vie{p}{\ellb}{},\Vie{p}{\mathbf{m}}{})
  = k_1 k_2 \begin{pmatrix}
  % pp
  \vecUnit{e}_{j,p}^b(\Vie{p}{\ellb}{})\cdot\vecUnit{e}_{j,p}^a (\Vie{p}{\mathbf{m}}{}) &
  % ps
  \vecUnit{e}_{k,p}^b(\Vie{p}{\ellb}{})\cdot\vecUnit{e}_{s}(\Vie{p}{\mathbf{m}}{}) \\
  % sp
  \vecUnit{e}_{s}(\Vie{p}{\ellb}{})\cdot\vecUnit{e}_{k,p}^a(\Vie{p}{\mathbf{m}}{}) &
  % ss
  \vecUnit{e}_{s}(\Vie{p}{\ellb}{})\cdot\vecUnit{e}_{s}(\Vie{p}{\mathbf{m}}{}) \\
  \end{pmatrix},
  \label{Mdef}
\end{equation}
\sloppy and it originates from a change of coordinate system between the local polarization basis  $(\hat{\mathbf{e}}_{j,p}^b (\Vie{p}{\ellb}{}),\hat{\mathbf{e}}_s (\Vie{p}{\ellb}{}))$ and $(\hat{\mathbf{e}}_{k,p}^a (\Vie{p}{\mathbf{m}}{}),\hat{\mathbf{e}}_s (\Vie{p}{\mathbf{m}}{}))$, defined for $a = \pm$, $b = \pm$, and $j, k \in \{1, 2\}$ with $j \neq k$.
The kernel scalar factor $\Cie{J}{jk}{b,a} (\Vie{p}{\ellb}{},\Vie{p}{\mathbf{m}}{})$ encodes the surface geometry and is defined as
\begin{equation}
\Cie{J}{jk}{b,a} (\Vie{p}{\ellb}{},\Vie{p}{\mathbf{m}}{}) = \Big[ b \alpha_j (\Vie{p}{\ellb}{}) - a \alpha_k (\Vie{p}{\mathbf{m}}{}) \Big]^{-1} \,  \frac{1}{a_c} \int_{a_c} \exp \Big[-i \big(\Vie{k}{j}{b} (\Vie{p}{\ellb}{}) - \Vie{k}{k}{a} (\Vie{p}{\mathbf{m}}{}) \big) \cdot \vec{s}(\pvec{x}) \Big] \dtwox ,
  \label{Iintdef}
\end{equation}
\end{widetext}
where $a_c$ denotes both the unit cell and its area and $\Vie{s}{}{} (\Vie{x}{\parallel}{}) = x_1 \, \Vie{\hat{e}}{1}{} + x_2 \, \Vie{\hat{e}}{2}{} + \zeta(\Vie{x}{\parallel}{}) \, \Vie{\hat{e}}{3}{}$ is a point on the surface. In the specific case of a one-dimensional sinusoidal profile of equation $x_3 = \zeta(\Vie{x}{\parallel}{}) =  H \, \sin(2 \pi x_1 / a)$, it is straightforward to show that~\cite{banon:thesis}
\begin{widetext}
\begin{equation}
\Cie{J}{jk}{b,a} (\Vie{p}{\ellb}{},\Vie{p}{\mathbf{m}}{}) = \frac{(-1)^{\ell_1 - m_1}}{\gamma_{jk}^{b,a}(\Vie{p}{\ellb}{},\Vie{p}{\mathbf{m}}{})} \, \delta_{\ell_2,m_2} \, J_{\ell_1 - m_1} \big[ \gamma_{jk}^{b,a}(\Vie{p}{\ellb}{},\Vie{p}{\mathbf{m}}{}) H \big] \: ,
\label{eq:Jsinus}
\end{equation}
\end{widetext}
where $J_n$ denotes the Bessel function of the first kind of order $n$ and we have introduced the short-hand notation $\gamma_{jk}^{b,a}(\Vie{p}{\ellb}{},\Vie{p}{\mathbf{m}}{}) \equiv b \alpha_j (\Vie{p}{\ellb}{}) - a \alpha_k (\Vie{p}{\mathbf{m}}{})$. Furthermore the reciprocal lattice vectors read $\Vie{G}{\ellb}{} = \frac{2\pi \ell_1}{a} \Vie{\hat{e}}{1}{}$ and all transmission amplitudes $\Vie{T}{}{(m_1,m_2)}$ vanish for $m_2 \neq 0$.  Consequently, Eq.~(\ref{RREint}) reduces to a similar system of equations but reduced to $m_2 = 0$. The reduced Rayleigh equation for 3 media takes a similar form but with modified kernel and right-hand side. Explicit expressions can be found in Refs.~\onlinecite{soubret1,Gonzalez-Alcalde:16,Banon:17:2,banon:thesis}.

\subsection{Numerical solution} 
In order to compute the diffraction efficiencies, we need to solve the reduced Rayleigh equation Eq.~(\ref{RREint}) to obtain the transmission amplitudes $T_{\mu \nu}^{(\ellb)}(\Vie{p}{0}{})$. Numerically, the infinite dimensional linear system is truncated to keep only a certain number of modes, $|\ellb| < N$. Assuming that the surface profile is smooth enough, the numerical solution of the truncated linear system converges towards the solution of the infinite dimensional system as $N \to \infty$. Numerically, we choose to stop increasing $N$ once the consecutive solutions do not change significantly.\\

\subsection{Single scattering approximation} 
A closed form approximate solution of Eq.~(\ref{RREint}) can be obtained as a Born approximation, which is accurate in the single scattering regime. Such a solution is derived in Appendix~\ref{appendix:born}.\\

\subsection{Diffraction efficiencies} 
It is experimentally challenging to access directly the complex-valued transmission amplitudes as one is often restricted to measuring intensities. For a periodic grating, the so-called diffraction efficiencies for each modes are often the observables of interest. The diffraction efficiency for the transmitted mode characterized by $(\mu, \ellb)$ scattered in medium $j$ given an incident plane wave characterized by $(\nu, \Vie{p}{0}{})$ is defined for propagating modes as
\begin{equation}
t_{\mu \nu}^{(\ellb)} (\Vie{p}{0}{}) \equiv \frac{\mathrm{Re} \, \left[ \alpha_j(\Vie{p}{\ellb}{}) \right] }{\alpha_1(\Vie{p}{0}{})} \, \Big| T_{\mu \nu}^{(\ellb)} (\Vie{p}{0}{}) \Big|^2  \: ,
\label{eq:efficiency}
\end{equation}
and $j = 2$ or $3$ depending on whether two or three media are considered. It represents the fraction of the incident flux of the incident plane wave characterized by the lateral wave vector $\Vie{p}{0}{}$ and polarization state $\nu$ which is transmitted into a plane wave in medium $j$ characterized by the lateral wave vector $\Vie{p}{\ellb}{}$ and polarization state $\mu$. For evanescent waves, the efficiency vanishes as $\alpha_j$ becomes pure imaginary. Experimentally, the source is a finite size beam and not a plane wave. The diffraction efficiencies can be estimated by integrating the power per unit solid angle and unit incident power over the solid angle defined by the considered diffraction peak. In our case, since the measurements have been restricted to the scattering plane and not the whole half space, we assume the efficiency to be proportional to the area under each peak of the BTDF cuts presented in Fig.~\ref{fig:btdf}. Hence we will compare experimentally and numerically the following \emph{normalized} transmission efficiencies
\begin{subequations}
\begin{align}
\tilde{t}_{\mu \nu}^{(\ell)} (\Vie{p}{0}{}) &\eqexp \frac{\int_{\theta_\ell - \Delta \theta}^{\theta_\ell + \Delta \theta} \mathrm{BTDF(\theta)} \cos(\theta) \: \mathrm{d}\theta}{\int_{-\pi / 2}^{\pi / 2} \mathrm{BTDF(\theta)} \cos(\theta) \: \mathrm{d}\theta} \: ,
\\
\tilde{t}_{\mu \nu}^{(\ell)} (\Vie{p}{0}{}) &\eqnume \frac{t_{\mu \nu}^{(\ell)} (\Vie{p}{0}{})}{\displaystyle \sum_{\ell \in \mathbb{Z}} t_{\mu \nu}^{(\ell)} (\Vie{p}{0}{})} \: . \label{eq:norm:eff}
\end{align}
\end{subequations}
Here the notation has been lighten to keep only $\ell = \ell_1$ instead of $(\ell_1,0)$ as only a cut for $\ell_2 = 0$ is assumed. The angular window $\Delta \theta$ is characteristic of the width of the diffraction peaks.

\subsection{Angular width of significant scattering}
The experimental BDTF curves shown in Fig.~\ref{fig:btdf} exhibit a angular window in which most of the diffracted power is concentrated (note the logarithmic scale). The typical angular width of significant scattering seems to shrink as the  surface profile relaxes and becomes smoother. To quantify this effect we define the angular width of significant scattering by
\begin{equation}
\sigma_\theta^2 = \sum_\ell \theta_\ell^2 \: \tilde{t}_{\mu \nu}^{(\ell)} (\Vie{p}{0}{}) - \Big( \sum_\ell \theta_\ell \: \tilde{t}_{\mu \nu}^{(\ell)} (\Vie{p}{0}{}) \Big)^2 \: .
\end{equation}
In other words, $\sigma_\theta$ is the standard deviation of the angles of diffraction where the normalized efficiencies play the role of a probability measure (note that by definition, we have indeed $\tilde{t}_{\mu \nu}^{(\ell)} (\Vie{p}{0}{}) \geq 0$ and $\sum_\ell \tilde{t}_{\mu \nu}^{(\ell)} (\Vie{p}{0}{}) = 1$). In principle, this quantity may depend on the polarization coupling $\mu$ and $\nu$ and the angle of incidence, but since the measurements are done with incident p-polarized light and at normal incidence we will not write explicitly this dependency in $\sigma_\theta$ for clarity. For the numerical results, the efficiencies are hence further averaged over polarization of the scattered field. A scaling law for $\sigma_\theta$ is derived in Appendix~\ref{appendix:angular_window} within the single scattering regime, for sufficiently smooth surface profiles and in the case of two media. It is of the form
\begin{equation}
\sigma_\theta = c_\zeta \: \left| \frac{n_2}{n_1} - 1 \right| \: \frac{2\pi H}{a} \: .
\label{eq:sigma:ss}
\end{equation}
The constant $c_\zeta$ depends on the shape of the profile [see Eq.~(\ref{eq:sigma:scaling})], and $n_1 = \sqrt{\varepsilon_1}$, $n_2 = \sqrt{\varepsilon_2}$ are the indices of refraction of the incident medium and of the sample material, respectively. The parameter $H$ denotes a characteristic amplitude of the surface profile and $a$ is the lattice constant. Physically, this scaling law asserts that the angular width of significant scattering, $\sigma_\theta$, increases linearly with a characteristic average slope $\frac{2\pi H}{a}$ of the profile. The rate of change of this linear behaviour is proportional to the index contrast $| n_2/n_1 - 1 |$, meaning that for a given surface profile, the width of significant scattering increases with the index contrast. Note that for $n_2 = n_1$, the index contrast vanishes and so does $\sigma_\theta$, as expected since there would not be any scattering of the light in such a case.

\subsection{Haze} 
Another quantity of interest in our study is the haze~\cite{Simonsen2002-1}. It quantifies the ratio of the power scattered in transmission at scattering angles outside an angular cone of $2.5^\circ$ around the specular direction to the total transmitted power. In the present context, the haze is thus related to the efficiencies by
\begin{equation}
\mathrm{Haze} = \sum_\ell \tilde{t}_{\mu \nu}^{(\ell)} (\Vie{p}{0}{}) \: \ind_{|\theta|> 2.5^\circ} (\theta_\ell - \theta_0) \: ,
\end{equation}
where $\ind_{|\theta|> 2.5^\circ}$ is the indicator function of angles larger than $2.5^\circ$.

% \clearpage
%-------------------------------------------------------------------------
\section{Results and discussion}\label{sec:results}
%-------------------------------------------------------------------------

\subsection{Main features of the diffraction patterns}
Let us analyze in some detail the diffraction patterns obtained experimentally. Several observations can be made from Fig.~\ref{fig:btdf}. First, the BTDF can be split in three angular regions. (i) An angular region in the neighborhood of the specular peak in which most of the scattered power is concentrated and makes a high intensity plateau (in a logarithmic scale). In this central region, a modulation of the intensity of the diffraction peaks is clearly present. These modulations and the width of this central region depend on the surface profile. (ii) Beyond the aforementioned central region, a fast exponential decay of the diffracted intensity occurs. The decay rate seems to be different for the different surface profiles. (iii) The exponential decay with the angle of diffraction saturates when the intensity reaches a low intensity plateau. Note that this intensity plateau is roughly the same for all samples, and does not seem to depend much on the surface profile. One could argue that we have reached the detector limit but this is not the case. The dynamic range of OMS4 setup is at least $10^7$, while signals as low as $8.10^{-3} sr^{-1}$ were measured by authors on mat absorbing samples~\cite{Colette_NCS} which is in a good agreement with the values reported for  black spectralon~\cite{Stover}. Also note that the diffraction peaks are still resolved at well defined diffraction angles. At the detector lower limit, the signal would be expected to be randomly noisy. The low intensity plateau may thus be the result of a scattering effect of the samples.\\

Considering now the sample with the smoothest profile (810 min. of annealing time), we compare the experimental normalized efficiencies, computed from the BTDF from Fig.~\ref{fig:btdf}, and the simulated efficiencies obtained by solving numerically the reduced Rayleigh equation. Two simulations were made: the first one assuming that the surface profile separated two semi-infinite homogeneous media (vacuum and a dielectric), and the second assuming a third medium (vacuum, dielectric, vacuum), i.e., taking into account the presence of the back surface of the glass. The simulation were made with the measured average surface profile displayed in Fig.~\ref{fig:profile} and with an index of refraction of the sample estimated to be $n_2 = 1.46$. For the simulation with three media, the mean thickness of the sample is taken consistently with the experimental thickness, $d = 2005.3~\mu$m. For the simulation with two media, the angles of diffraction in the dielectric are of course different than in the vacuum, but we simply keep the corresponding efficiencies and plot them as a function of the order index rather than the angle of diffraction. A Fresnel transmission from the dielectric to vacuum could, in principle,  be made as a correction but we have chosen not to do so. We have experienced that it does not change much the results and it is not needed as the simulation with three media was available.

%-------------------------------------------------------------------------
% --- Figure
%-------------------------------------------------------------------------
\begin{figure}[t]
\centering
\includegraphics[width = 0.45\textwidth]{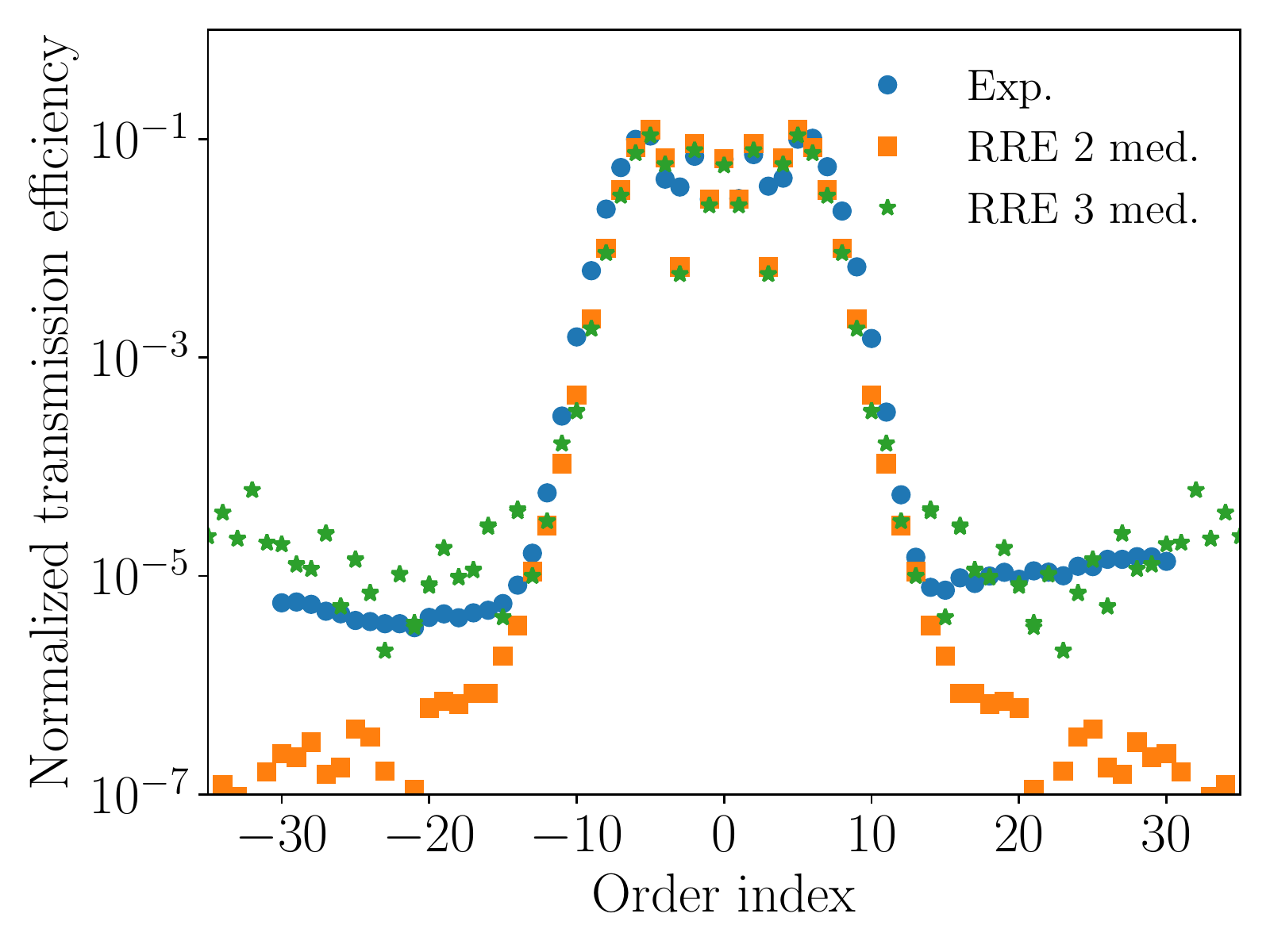}
\caption{Normalized transmission efficiency as a function of the diffractive order index for a normally incident p-polarized plane wave with wavelength in vacuum $\lambda = 520$~nm. The grating profile used for the simulation is the profile measured for the smoothest sample (810 minutes relaxation time). Blue circles: experimental data; Orange squares: simulation data obtained with the reduced Rayleigh equation solver assuming two media $n_1 = 1$ and $n_2 = 1.46$; Green stars: simulation data obtained with the reduced Rayleigh equation solver assuming three media $n_1 = 1$ and $n_2 = 1.46$ and $n_3 = 1$. In the latter case, the mean thickness of medium 2 is assumed to be $d = 2005.3~\mu$m in agreement with experimental estimates.}
\label{fig:3}
\end{figure}
%-------------------------------------------------------------------------

%-------------------------------------------------------------------------
% --- Figure
%-------------------------------------------------------------------------
\begin{figure}[t]
\centering
\includegraphics[width = 0.45\textwidth]{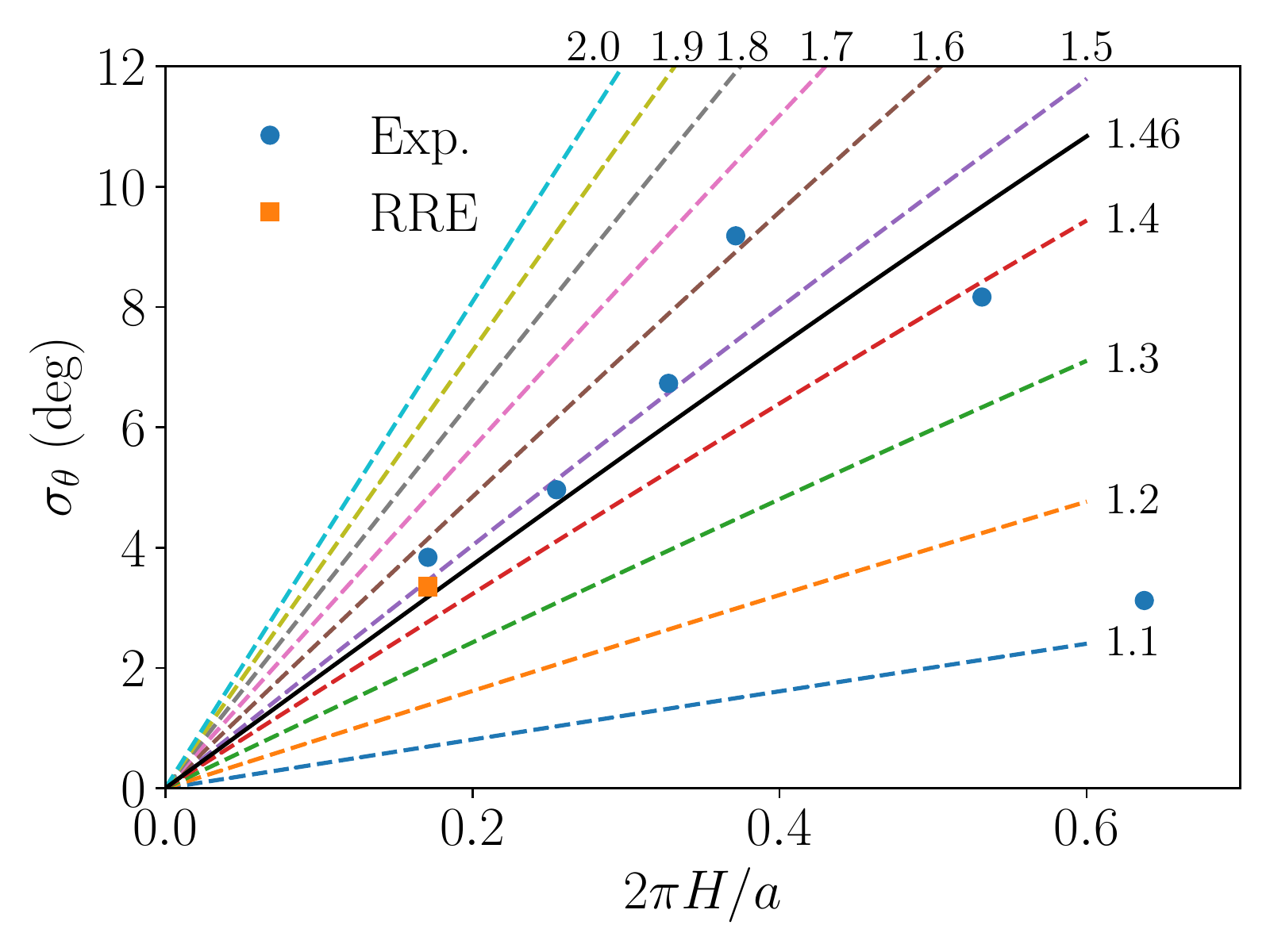}
\caption{Angular width of significant diffraction as a function of $2 \pi H / a$. Blue circles: experimental data; Orange square: simulation data obtained with the reduced Rayleigh equation solver assuming two media $n_1 = 1$ and $n_2 = 1.46$. Only the smoothest experimental profile could be simulated satisfactorily (810 minutes relaxation time). The dashed lines correspond to results obtained with reduced Rayleigh equation solver for \emph{sinusoidal} profiles and various refractive index. The value of the refractive index is indicated by the end of the line. The black solid line indicate the simulation results for $n_2 = 1.46$.}
\label{fig:4}
\end{figure}
%-------------------------------------------------------------------------

 The experimental and simulated normalized transmission efficiencies are shown in Fig.~\ref{fig:3}. Although the individual simulated diffraction efficiencies do not match perfectly the experimental ones, we can see that the simulated data reproduce the main features observed experimentally. In particular, the relative order of magnitudes of efficiency are rather well reproduced in the three aforementioned angular regions. To be more accurate, both the simulation with two media and with three media, reproduce well the overall level of intensity in the central region, the width of the central region, and the exponential decay observed experimentally. The two simulations then deviate from one another for large diffraction orders, in the low efficiency plateau. The simulation with three media correctly catches the low intensity plateau while the simulation with two media continues with a exponential decay to much lower efficiency level. We interpret this result as follows. The large efficiencies are essentially the result of single scattering events (this hypothesis seems indeed verified numerically, although not shown here) which explains that the two simulations are relatively close in the central and decay regions. The single scattering approximation of the transmission efficiency for a one-dimensional sinusoidal surface is shown to be proportional to (see Appendix~\ref{appendix:born})
\begin{equation}
T^{(\ellb)}_{\mu \nu} (\Vie{p}{0}{}) \propto J_{\ell_1} \Big( \gamma_{12}^{-,-}(\Vie{p}{\ellb}{},\Vie{p}{0}{}) H \Big) \: ,
\end{equation}
where $J_\ell$ is the Bessel function of the first kind of order $\ell$. The single scattering approximation in fact predicts the modulation of the diffraction efficiencies in the central angular region and then an exponential decay without lower bound with increasing diffraction order index. This being true for both configurations (2 or 3 media). The low efficiency plateau thus results from contribution from multiple scattering. We speculate that in the three media configuration, the optical paths multiply reflected between the two interfaces of the sample increase the multiple scattering level since the optical paths have more chances of being scattered on the structured surface.\\

\subsection{Angular width of significant scattering and haze} 
These observations being made, we will now focus our attention on integrated observables such as the angular width of significant scattering, $\sigma_\theta$, and the haze. Since we have shown that, at least for the smoothest profile, the theory with two media is sufficient to explain the dominant contribution to the efficiencies, we now only consider this theory to simplify the analysis.

 Figure~\ref{fig:4} shows the angular width of significant scattering plotted as a function of $2\pi H/a$, where $a$ is the grating period and $H$ is half the maximum amplitude of the surface profile (i.e. $2H = \max \zeta -\min \zeta $). The experimental data points are plotted together with the simulation achieved for the smoothest experimental profile. For the remaining profiles, the simulations were unstable due to numerical error in the evaluation of the integral kernel $\Cie{J}{12}{-,-}$ which becomes challenging to evaluate as $H / \lambda$ grows. Fortunately, the integral kernel can be evaluated accurately numerically thanks to the analytical expression in terms of Bessel functions for sinusoidal profiles. Simulation for sinusoidal profiles with larger profile amplitudes could then be achieved and the results are represented continuously by the lines in Fig.~\ref{fig:4}. The different dashed lines are computed by solving numerically the reduced Rayleigh equation for a sinusoidal profile for different values of the refractive index of the substrate, thus drawing an abacus. The black solid line corresponds to the results obtained with a refractive index of $n_2=1.46$ in agreement with the experimental estimate.

First, let us observe that for sinusoidal profiles the simulated $\sigma_\theta$ results, which \emph{do not} assume the single scattering approximation, behave linearly with $2 \pi H / a$ in a strikingly accurate way. This is in agreement with the single scattering scaling law, Eq.~(\ref{eq:sigma:ss}), derived in Appendix~\ref{appendix:angular_window}. It therefore comforts us in the single scattering interpretation which we have motivated earlier. Second, we also observe the increase of the slope, $\mathrm{d} \sigma_\theta / \mathrm{d} (2 \pi H / a)$, with increasing refractive index, as predicted by the factor $| n_2/n_1 - 1 |$ in Eq.~(\ref{eq:sigma:ss}). Figure~\ref{fig:dsigma} in Appendix shows the excellent agreement in the variation of the slope with $| n_2/n_1 - 1 |$ as predicted from the analysis (the constant $c_\zeta = 1 / \sqrt{2}$ being analytically known for sinusoidal profile). Turning now, to the comparison with the experimental data points, we can see that experimental data points overall agree relatively well with the solid line computed for sinusoidal profiles, at least for the smoothest profiles. The data point for the steepest profile deviates significantly from the prediction. This is expected to occur since a sinusoidal approximation of the surface profile become erroneous.

\begin{figure}[t]
\centering
\includegraphics[width = 0.45\textwidth]{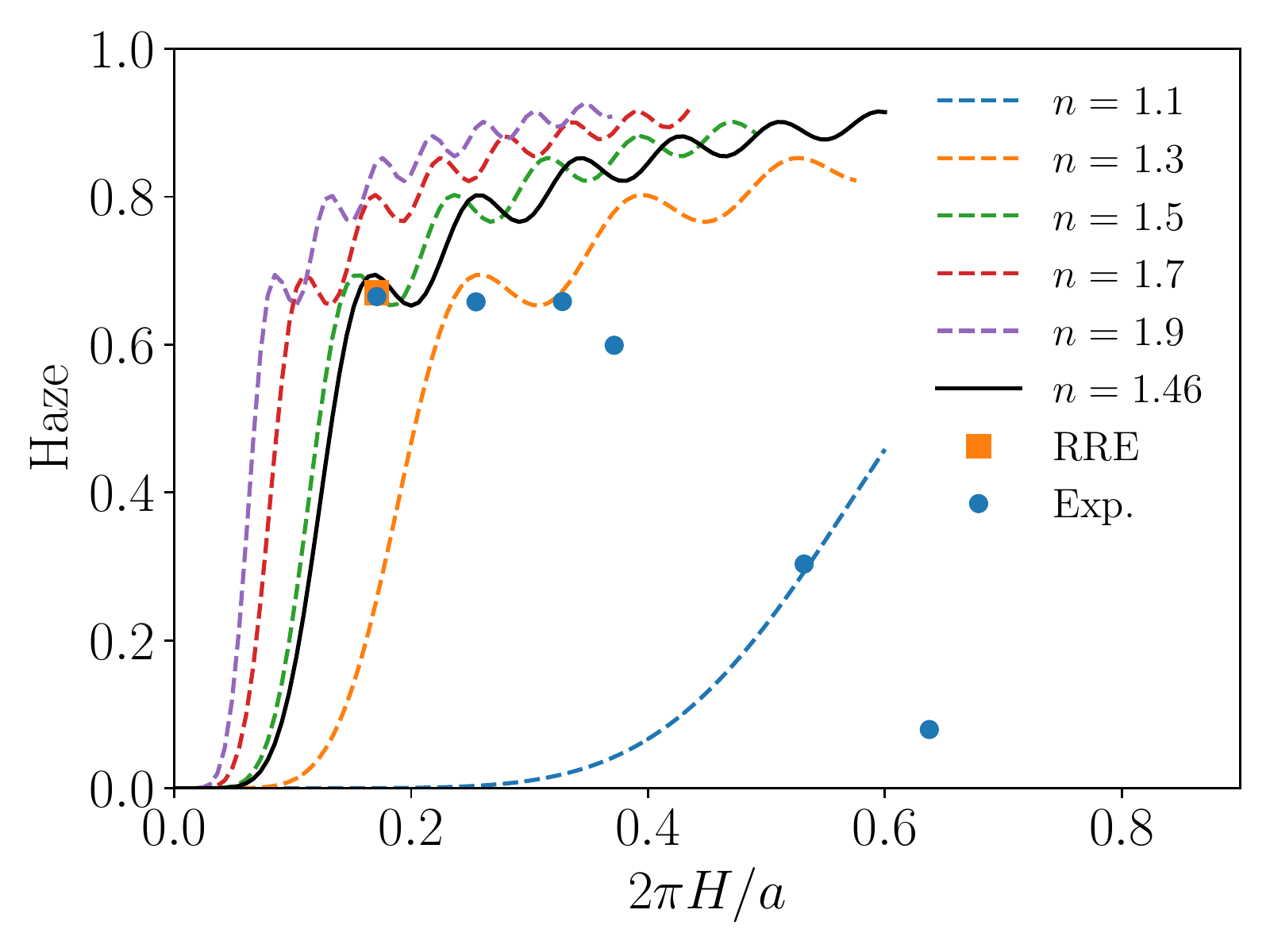}
\caption{Haze as a function of $2 \pi H / a$. Blue circles: experimental data; Orange square: simulation data obtained with the reduced Rayleigh equation solver assuming two media $n_1 = 1$ and $n_2 = 1.46$. Only the smoothest experimental profile could be simulated satisfactorily (810 minutes). The dashed lines correspond to results obtained with reduced Rayleigh equation solver for \emph{sinusoidal} profiles and various refractive index. The black solid line indicate the simulation results for $n = 1.46$.}
\label{fig:5}
\end{figure}

Let us now consider the haze as a function of $2 \pi H / a$ shown on Fig.~\ref{fig:5}. The simulated haze for sinusoidal profiles exhibits an overall sigmoid shape. For low profile amplitudes, the haze is essentially zero, and as the profile amplitude increases, the haze rises quickly to a value around 60 to 70\% and then grows with an oscillatory behavior towards unity. The overall shape is explained as follows. For a planar surface, the haze is zero since all the transmitted power goes into the specular direction. As the profile amplitude is increased, the haze remains close to zero until the angular window of significant scattering reaches the angular cone of 2.5$^\circ$, acting as a threshold. Then the haze grows linearly with the linear growth of $\sigma_\theta$. Since the haze is by definition bounded by one, it cannot, of course, grow linearly indefinitely. Some kind of saturation occurs, and the oscillations are the signature of the modulation of the efficiency as the profile amplitude is changed, and as the transmitted power is redistributed between the diffraction orders inside and outside the \ang{2.5} cone.

 The haze data point corresponding to the smoothest profile agrees remarkably well with the prediction, while larger deviations are observed for the remaining profiles. The four smoothest profiles have similar values of haze around $60 - 65 \%$ while the two last exhibit significantly lower values around $30\%$ and $8 \%$, respectively. This haze drop may seem surprising at first. Indeed, one may expect that the steepest the profile, the broader the diffraction pattern. This picture is also somewhat illustrated in Fig.~\ref{fig:btdf} showing the BTDF for the two steepest profiles (0 and 30 min. annealing time). However, the same curves also show that the specular peak has a much larger intensity than the remaining orders. Even though the BTDF is wide the amount of power going into the large angles of diffraction is relatively low compared to the power going into the specular peak. The careful observation of the BTDF explains the decay of the haze. But what is the physical reason for this re-concentration of the power into the specular direction as the profile becomes steep? By inspection of the steepest profile in Fig.~\ref{fig:btdf}, we see that it could be well approximated by a square profile, where the bottom and the top of the profile would exhibit planar sections. Consequently, it is not unreasonable to think that in this limit, this sequence of planar sections tends to transmit more in the specular direction, which would explain the decay of the haze value. Note that there is also a signature of this deviation in the corresponding data point for $\sigma_\theta$ in Fig.~\ref{fig:4}, which is experimentally lower than the predicted value for a sinusoidal profile.

%-------------------------------------------------------------------------
% --- Figure
%-------------------------------------------------------------------------
%-------------------------------------------------------------------------

% Conclusion
%-------------------------------------------------------------------------
\section{Conclusion}\label{sec:conclusion}
%-------------------------------------------------------------------------

Our experimental, numerical and analytical study has allowed us to better understand the behavior of the diffraction properties of large period dielectric gratings at different levels of detail. First, we have understood the different angular regions of efficiency level and the modulation of the efficiencies with the angle of diffraction based on a single scattering approach, and taking into account or not the finite thickness of the sample.
Furthermore, we have derived a simple scaling law for the angular width of significant scattering which is well satisfied experimentally for sufficiently smooth profiles. In addition, we have observed deviation to this law for steep profiles, and suggested a physically realistic explanation of re-concentration into the specular direction as a consequence of the profile becoming similar to a square profile, i.e., a sequence of planar interfaces. This speculated explanation would explain the lowering of the width of significant scattering and of the haze values.

The tools developed in this work may be useful for the fast characterization of dielectric gratings, in particular in obtaining at a low computational cost critical dimensions such as the grating period by using the grating formula, and the profile amplitude by using the angular width of significant scattering or the haze. The scaling law Eq.~(\ref{eq:sigma:ss}) could also help as a simple guide for designing patterns with desired angular scattering width for the purpose of visual appearance for which one may not be so exigent on the individual diffraction efficiencies but more on integrated quantities over angular windows. Nevertheless, we have seen that other tools will be needed for the characterization of steeper profiles, in particular with the use of more adapted numerical methods (cf. Ref.~\onlinecite{Petit:1980}) than the one used in the present work, but at a higher computational cost considering the large period to wavelength ratio of the type of gratings studied here.

% \begin{acknowledgments}
% --------------------------------------------------------------------
\section*{Acknowledgment}
% --------------------------------------------------------------------

We thank Jeremie Teisseire for valuable discussions and Emmanuel Garre for invaluable help with the fabrication of samples.
This research was supported by the French National Research Agency (ANR-15-CHIN-0003) and the Simons foundation Grant No. 601944.
% \end{acknowledgments}

\appendix
%-------------------------------------------------------------------------
\section{Rayleigh equations of the second kind and Born approximation} \label{appendix:born}
%-------------------------------------------------------------------------
The numerical solution of the reduced Rayleigh equation gives accurate estimates of the scattering amplitudes for smooth enough surfaces but may require to set up and solve a somewhat large linear system. For a one-dimensional profile under non-conical incidence, the system size grows linearly as $\mathcal{O}(N)$ and does not represent a real issue. Nevertheless, for bi-periodic gratings the size of the system grows as $\mathcal{O}(N^2)$ and the use of approximate closed form solutions may be of particular interest in order to derive scaling laws relating parameters of the surface profile and features of the scattering data for sample characterization. We present here such an approximate solution, which is the Born approximation. It is constructed from the reduced Rayleigh equation of the \emph{second kind} which derives from the "classic" reduced Rayleigh equation. We derive this approximate solution for the transmission amplitudes, but the same procedure applies for reflection (see Refs.~\onlinecite{banon:thesis,Maradudin:83}).\\

Consider the reduced Rayleigh equation, Eq.~(\ref{RREint}), for the transmission amplitudes to which we apply two steps. First, we expand the scalar kernel factor $\Cie{J}{12}{-,-}$ as the sum of two term:
\begin{equation}
\Cie{J}{12}{-,-} (\Vie{p}{\ellb}{},\Vie{p}{\mathbf{m}}{}) = \frac{\delta_{\ellb,\mathbf{m}}}{\gamma_{12}^{-,-}(\Vie{p}{\ellb}{},\Vie{p}{\mathbf{m}}{})} + \Cie{K}{12}{-,-} (\Vie{p}{\ellb}{},\Vie{p}{\mathbf{m}}{}) \: ,
\label{eq:kernel:splitting}
\end{equation}
which defines $\Cie{K}{12}{-,-} (\Vie{p}{\ellb}{},\Vie{p}{\mathbf{m}}{})$ as
\begin{widetext}
\begin{equation}
\Cie{K}{12}{-,-} (\Vie{p}{\ellb}{},\Vie{p}{\mathbf{m}}{}) = \frac{1}{\gamma_{12}^{-,-}(\Vie{p}{\ellb}{},\Vie{p}{\mathbf{m}}{})} \: \frac{1}{a_c} \int_{a_c} \exp \Big[-i(\Vie{p}{\ellb}{} - \Vie{p}{\mathbf{m}}{}) \cdot \pvec{x} \Big] \:  \Big\{ \exp \Big[- i \gamma_{12}^{-,-}(\Vie{p}{\ellb}{},\Vie{p}{\mathbf{m}}{}) \, \zeta(\pvec{x}) \Big] - 1 \Big\} \: \dtwox .
  \label{Kintdef}
\end{equation}
\end{widetext}
The first term on the right-hand side of Eq.~(\ref{eq:kernel:splitting}) corresponds to what $\Cie{J}{12}{-,-} (\Vie{p}{\ellb}{},\Vie{p}{\mathbf{m}}{})$ reduces to in the case of a planar interface, i.e. $\zeta = 0$, and the second terms hence corresponds to the deviation of the scalar kernel factor from this reference value.
Second, we introduce the following change of unknown. We write the unknown transmission amplitude as the sum of the transmission amplitude we would obtain for a planar interface (the Fresnel transmission amplitude), $\boldsymbol \tau_0 (\Vie{p}{0}{}) \, \delta_{\ellb,\mathbf{0}}$, and a remaining unknown term, $\Delta \Vie{T}{}{(\ellb)}(\Vie{p}{0}{})$, to be determined, and which corresponds to the deviation of the transmission amplitude from the Fresnel amplitude
\begin{widetext}
\begin{subequations}
\begin{align}
\boldsymbol \tau_0 (\Vie{p}{0}{})  &= \frac{2 n_1 n_2 \, \alpha_1(\Vie{p}{0}{})}{\epsilon_2 -\epsilon_1} \big[\alpha_2(\Vie{p}{0}{}) - \alpha_1(\Vie{p}{0}{}) \big] \: \Big[ \Vie{M}{12}{-,-}(\Vie{p}{0}{},\Vie{p}{0}{}) \Big]^{-1} \: , \\
\Vie{T}{}{(\ellb)}(\Vie{p}{0}{}) &= \boldsymbol \tau_0 (\Vie{p}{0}{}) \, \delta_{\ellb,\mathbf{0}} + \Delta \Vie{T}{}{(\ellb)}(\Vie{p}{0}{}) \: .
\end{align}
\label{eq:T:splitting}
\end{subequations}
By plugging Eqs.~(\ref{eq:kernel:splitting}) and (\ref{eq:T:splitting}) into Eq.~(\ref{RREint}) for the transmission amplitude we obtain after simplification of the planar interface contribution the following equation for $\Delta \Vie{T}{}{(\ellb)}(\Vie{p}{0}{})$
\begin{align}
&\big[ \alpha_2(\Vie{p}{\ellb}{}) - \alpha_1(\Vie{p}{\ellb}{}) \big]^{-1} \, \Vie{M}{12}{-,-}(\Vie{p}{\ellb}{},\Vie{p}{\ellb}{}) \: \Delta \Vie{T}{}{(\ellb)}(\Vie{p}{0}{}) \nonumber \\
&+ \sum_{\mathbf{m} \in \mathbb{Z}^2} \Cie{K}{12}{-,-} (\Vie{p}{\ellb}{},\Vie{p}{\mathbf{m}}{}) \, \Vie{M}{12}{-,-}(\Vie{p}{\ellb}{},\Vie{p}{\mathbf{m}}{}) \: \Delta \Vie{T}{}{(\mathbf{m})}(\Vie{p}{0}{}) = - \Cie{K}{12}{-,-} (\Vie{p}{\ellb}{},\Vie{p}{0}{}) \, \Vie{M}{12}{-,-}(\Vie{p}{\ellb}{},\Vie{p}{0}{}) \: \boldsymbol \tau_0 (\Vie{p}{0}{}) \: .
\label{eq:RRE2}
\end{align}
Equation~(\ref{eq:RRE2}) is a linear system which resembles a Fredholm integral equation of the second kind and can be solved efficiently by successive iteration of a fixed point algorithm (assuming convergence). The first iterate, known as the Born approximation, is obtained by setting $\Delta \Vie{T}{}{(\mathbf{m})}(\Vie{p}{0}{}) = 0$ in the sum over $\mathbf{m}$ and reads
\begin{subequations}
\begin{align}
\Delta \Vie{T}{}{(\ellb)}(\Vie{p}{0}{}) \approx  &\Cie{K}{12}{-,-} (\Vie{p}{\ellb}{},\Vie{p}{0}{}) \,  \boldsymbol \tau (\Vie{p}{\ellb}{} ,\Vie{p}{0}{}) \\
\boldsymbol \tau (\Vie{p}{\ellb}{} ,\Vie{p}{0}{}) \equiv &\big[ \alpha_1(\Vie{p}{\ellb}{}) - \alpha_2(\Vie{p}{\ellb}{}) \big] \, \Big[ \Vie{M}{12}{-,-}(\Vie{p}{\ellb}{},\Vie{p}{\ellb}{}) \Big]^{-1}  \Vie{M}{12}{-,-}(\Vie{p}{\ellb}{},\Vie{p}{0}{}) \: \boldsymbol \tau_0 (\Vie{p}{0}{}) \: .
\label{eq:born}
\end{align}
\end{subequations}
This is a single scattering approximation of the solution of the reduced Rayleigh equation.\\
\end{widetext}

\subsection{One-dimensional sinusoidal profile}
In the specific case of a one-dimensional sinusoidal profile, the factor $\Cie{K}{12}{-,-} (\Vie{p}{\ellb}{},\Vie{p}{0}{})$ is known analytically since $\Cie{J}{12}{-,-} (\Vie{p}{\ellb}{},\Vie{p}{0}{})$ is given by Eq.~(\ref{eq:Jsinus}). In particular, for $\ell_1 \neq 0$ and $\ell_2 = 0$, we have
\begin{widetext}
\begin{equation}
\Vie{T}{}{(\ellb)}(\Vie{p}{0}{}) \approx \Delta \Vie{T}{}{(\ellb)}(\Vie{p}{0}{}) \approx \frac{(-1)^{\ell_1}}{\gamma_{12}^{-,-} (\Vie{p}{\ellb}{},\Vie{p}{0}{})} \: J_{\ell_1} \Big[ \gamma_{12}^{-,-} (\Vie{p}{\ellb}{},\Vie{p}{0}{}) H \Big] \boldsymbol \tau (\Vie{p}{\ellb}{} ,\Vie{p}{0}{})  \: .
\label{eq:Born:sinus}
\end{equation}
\end{widetext}

\section{Angular width of significant scattering} \label{appendix:angular_window}

We derive here a simple scaling law for the angular width of significant scattering $\sigma_\theta$ as a function of the  characteristic lengths of the surface profile. This law is derived in the single scattering regime and for small profile amplitude compared to the wavelength of the incident light.

 Instead of working directly with angles, we work with in-plane wave vectors. The angular width and the wave vector width of significant scattering being simply proportional in the regime of small angle of diffraction, which is what we assume here for simplicity. We will assume normal incidence $\Vie{p}{0}{} = \Vie{0}{}{}$, and a one-dimensional profile as this is the case needed in the present work. Hence we wish to estimate
\begin{equation}
\sigma_{\mu \nu}^2 = \sum_{\ell = - L}^{L}  |\Vie{p}{\ell}{}|^2 \: \tilde{t}_{\mu \nu}^{(\ell)} \: ,
\end{equation}
where $L = \lfloor \frac{n_1 \, a}{\lambda} \rfloor$ is the maximal index for which a diffracted mode propagates in vacuum and $\tilde{t}_{\mu \nu}^{(\ell)}$ are the normalized transmission efficiencies. By using that $|\Vie{p}{\ell}{}|^2 = (2 \pi \ell)^2 / a^2$ and the definition of $\tilde{t}_{\mu \nu}^{(\ell)}$ [Eq.~(\ref{eq:norm:eff})], the above equation can be recast as
\begin{equation}
\sigma_{\mu \nu}^2 = \frac{2 }{k_1 \mathcal{T}} \: \left( \frac{2 \pi}{a} \right)^2 \: \sum_{\ell = 1}^{L} \ell^2 \: \alpha_2 (\Vie{G}{\ell}{}) \: \big| T_{\mu \nu}^{(\ell)} \big|^2 \: ,
\label{eq:sigma}
\end{equation}
\sloppy where we have introduced the transmittance $\mathcal{T} = \sum_{\ell \in \mathbb{Z}} t_{\mu \nu}^{(\ell)}$. Furthermore, we have assumed that the profile is symmetric and hence, under normal incidence, $\big| T_{\mu \nu}^{(-\ell)} \big|^2 = \big| T_{\mu \nu}^{(\ell)} \big|^2$, and the term $\ell = 0$ does not contribute to the sum. We assume now that the surface profile is smooth enough such that $T_{\mu \nu}^{(\ell)}$ can be reasonably approximated by $\Delta T_{\mu \nu}^{(\ell)}$ given by the Born approximation Eq.~(\ref{eq:born}).

 For smooth surfaces, the significant scattering efficiencies occur in a small wave vector region around the specular direction. Hence $\alpha_2 (\Vie{G}{\ell}{}) \approx k_2$ for small $|\ell|$. This approximation will be crude for large $|\ell|$ but since $|T_{\mu \nu}^{(\ell)}|$ decays exponentially for large $|\ell|$ the approximation can be made for all $\ell$ without significantly changing the result. This yields
\begin{equation}
\sigma_{\mu \nu}^2 \approx \frac{2 \, k_2}{k_1 \mathcal{T}} \: \left( \frac{2 \pi}{a} \right)^2 \: \sum_{\ell = 1}^{L} \ell^2 \: \big| \Delta T_{\mu \nu}^{(\ell)} \big|^2 \: ,
\label{eq:sigma_intermediary}
\end{equation}
 Then comes the estimate of $\big| \Delta T_{\mu \nu}^{(\ell)} \big|^2$ which is proportional to $|\Cie{K}{12}{-,-} (\Vie{G}{\ell}{},\Vie{0}{}{}) |^2$ in the single scattering regime [Eq.~(\ref{eq:born})]. We have
%
%\begin{align}
%&|\Cie{K}{12}{-,-} (\Vie{G}{\ell}{},\Vie{0}{}{}) |^2 = a^{-2} \, |\gamma_{12}^{-,-}(\Vie{G}{\ell}{},\Vie{0}{}{})|^{-2} \nonumber \\
%&\int_{-a/2}^{a/2} \int_{-a/2}^{a/2} e^{- i  \frac{2 \pi \ell}{a} \, (x_1-x_1^\prime)} \: \Big[e^{- i \gamma_{12}^{-,-}(\Vie{G}{\ell}{},\Vie{0}{}{}) \zeta(x_1)} - 1 \Big] \nonumber \\
%&\times\Big[e^{ i \gamma_{12}^{-,-}(\Vie{G}{\ell}{},\Vie{0}{}{})^* \zeta(x_1^\prime)} - 1 \Big] \: \mathrm{d}x_1 \mathrm{d}x_1^\prime \: .
%\end{align}
\begin{equation}
    \Cie{K}{12}{-,-} (\Vie{G}{\ell}{},\Vie{0}{}{}) = \frac{\int_{-a/2}^{a/2} e^{- i  \frac{2 \pi \ell}{a} \, x_1} \: \Big[e^{- i \gamma_{12}^{-,-}(\Vie{G}{\ell}{},\Vie{0}{}{}) \zeta(x_1)} - 1 \Big]  \: \mathrm{d}x_1  }{a \gamma_{12}^{-,-}(\Vie{G}{\ell}{},\Vie{0}{}{})} \: .
\end{equation}
For surface profiles with amplitudes small compared with the wavelength, the factors in the square brackets can be linearly approximated, which leads to
%
%\begin{align}
%&|\Cie{K}{12}{-,-} (\Vie{G}{\ell}{},\Vie{0}{}{}) |^2 \approx a^{-2} \, \nonumber \\
%&\int_{-a/2}^{a/2} \int_{-a/2}^{a/2} e^{- i  \frac{2 \pi \ell}{a} \, (x_1-x_1^\prime)} \: \zeta(x_1) \zeta(x_1^\prime) \: \mathrm{d}x_1 \mathrm{d}x_1^\prime \: .
%\end{align}
\begin{equation}
    \Cie{K}{12}{-,-} (\Vie{G}{\ell}{},\Vie{0}{}{}) = \frac{1}{a} \int_{-a/2}^{a/2} e^{- i  \frac{2 \pi \ell}{a} \, x_1} \: \zeta(x_1) \: \mathrm{d}x_1 \: .
\end{equation}
By using a change of variables, $x_1 = a u$, and by writing $\zeta(x_1) = H \bar{\zeta}(y_1)$ where $H$ is a characteristic profile amplitude, we obtain
\begin{equation}
\Cie{K}{12}{-,-} (\Vie{G}{\ell}{},\Vie{0}{}{})  \approx H \, \Cie{F}{\zeta}{(\ell)} \: ,
\label{eq:approxK}
\end{equation}
with the dimensionless factor $\Cie{F}{\zeta}{(\ell)} $ defined by
\begin{equation}
\Cie{F}{\zeta}{(\ell)} = \int_{-1/2}^{1/2} e^{- i  2 \pi \ell \, u} \: \bar{\zeta}(u) \: \mathrm{d}u \: .
\end{equation}
By inserting Eq.~(\ref{eq:approxK}) into Eq.~(\ref{eq:sigma_intermediary}) we obtain
\begin{equation}
\sigma_{\mu \nu}^2 \approx \frac{2 \, k_2}{k_1 \mathcal{T}} \: \left( \frac{2 \pi H}{a} \right)^2 \: \sum_{\ell = 1}^{L} \ell^2 \, \big| \Cie{F}{\zeta}{(\ell)}\big|^2 \, \big| \tau_{\mu \nu} (\Vie{G}{\ellb}{} ,\Vie{0}{}{}) \big|^2 \: .
\end{equation}
From the definition of $\boldsymbol \tau$ [Eq.~(\ref{eq:born})], and together with our previous approximation of small angles of diffraction, we have
\begin{align}
   \boldsymbol \tau (\Vie{G}{\ellb}{} ,\Vie{0}{}{}) \approx \big( k_1 - k_2 \big) \, \Big[ \Vie{M}{12}{-,-}(\Vie{G}{\ellb}{},\Vie{G}{\ellb}{}) \Big]^{-1} \nonumber \\  \times \Vie{M}{12}{-,-}(\Vie{G}{\ellb}{},\Vie{0}{}{}) \: \boldsymbol \tau_0 (\Vie{p}{0}{}) \: ,
\end{align}
where for small angle of scattering the matrix product %
\begin{equation}
    \Big[\Vie{M}{12}{-,-}(\Vie{G}{\ellb}{},\Vie{G}{\ellb}{}) \Big]^{-1} \Vie{M}{12}{-,-}(\Vie{G}{\ellb}{},\Vie{0}{}{}) \approx \Vie{I}{2}{} \: ,
\end{equation}
and depends only weakly on the indices $n_1$, $n_2$. 
Consequently, we have
\begin{equation}
\sigma_{\mu \nu}^2 \approx  \big(k_1 - k_2 \big)^2  \: \left( \frac{2 \pi H}{a} \right)^2 \: g_{\zeta} \: \frac{k_2 \big|\tau_{\mu \nu, 0}\big|^2}{k_1 \mathcal{T}}  \: ,
\label{eq:transmittance:ratio}
\end{equation}
where $g_{\zeta}$ is given by
\begin{equation}
g_{\zeta}  =  2 \sum_{\ell = 1}^{L} \ell^2 \, \big| \Cie{F}{\zeta}{(\ell)}\big|^2 \: .
\end{equation}
The factor $g_{\zeta}$ depends only on the shape of the profile in the small angle of diffraction limit.
 The factor $k_2 \big|\tau_{\mu \nu, 0}\big|^2 / k_1 \mathcal{T}$ in Eq.~(\ref{eq:transmittance:ratio}) is simply the ratio between the transmittance of a planar surface and the transmittance of the patterned surface. Provided we are in the weakly scattering regime, these two transmittances are very close, $k_2 \big|\tau_{\mu \nu, 0}\big|^2 / k_1 \mathcal{T} \approx 1$, hence
 \begin{equation}
\sigma_{\mu \nu}^2 \approx  \big(k_1 - k_2 \big)^2  \: \left( \frac{2 \pi H}{a} \right)^2 \: g_{\zeta} \: .
\end{equation}
Finally, since $\theta^{(\ell)} = \arcsin \Big( |\Vie{G}{\ell}{}| / k_1 \Big) \approx |\Vie{G}{\ell}{}| / k_1$ in the considered regime, we deduce $\sigma_{\theta, \mu \nu}^2 \approx \sigma_{\mu \nu}^2 / k_1^2$. The angular width of significant scattering finally reads
\begin{equation}
\sigma_{\theta, \mu \nu} \approx g_{\zeta}^{1/2} \left| \frac{n_2}{n_1} - 1 \right| \:  \frac{2 \pi H}{a}   \: .
\label{eq:sigma:scaling}
\end{equation}
Note that in the special case of a sinusoidal profile $\zeta(x_1) = H \sin(2 \pi x_1 / a)$, we get
\begin{equation}
    \Cie{F}{\mathrm{sin}}{(\ell)} = \frac{\mathrm{sgn}(\ell) }{2i} \delta_{|\ell|,1} \: ,
\end{equation}
hence
\begin{equation}
    g_\mathrm{sin} = 1 / 2 \: ,
\end{equation}
and
\begin{equation}
\sigma_{\theta, \mathrm{sin}} \approx \frac{1}{\sqrt{2}} \left| \frac{n_2}{n_1} - 1 \right| \:  \frac{2 \pi H}{a}   \: .
\label{eq:sigma:scaling:sin}
\end{equation}
%
%-------------------------------------------------------------------------
% --- Figure
%-------------------------------------------------------------------------
\begin{figure}[t]
\centering
\includegraphics[width = 0.45\textwidth]{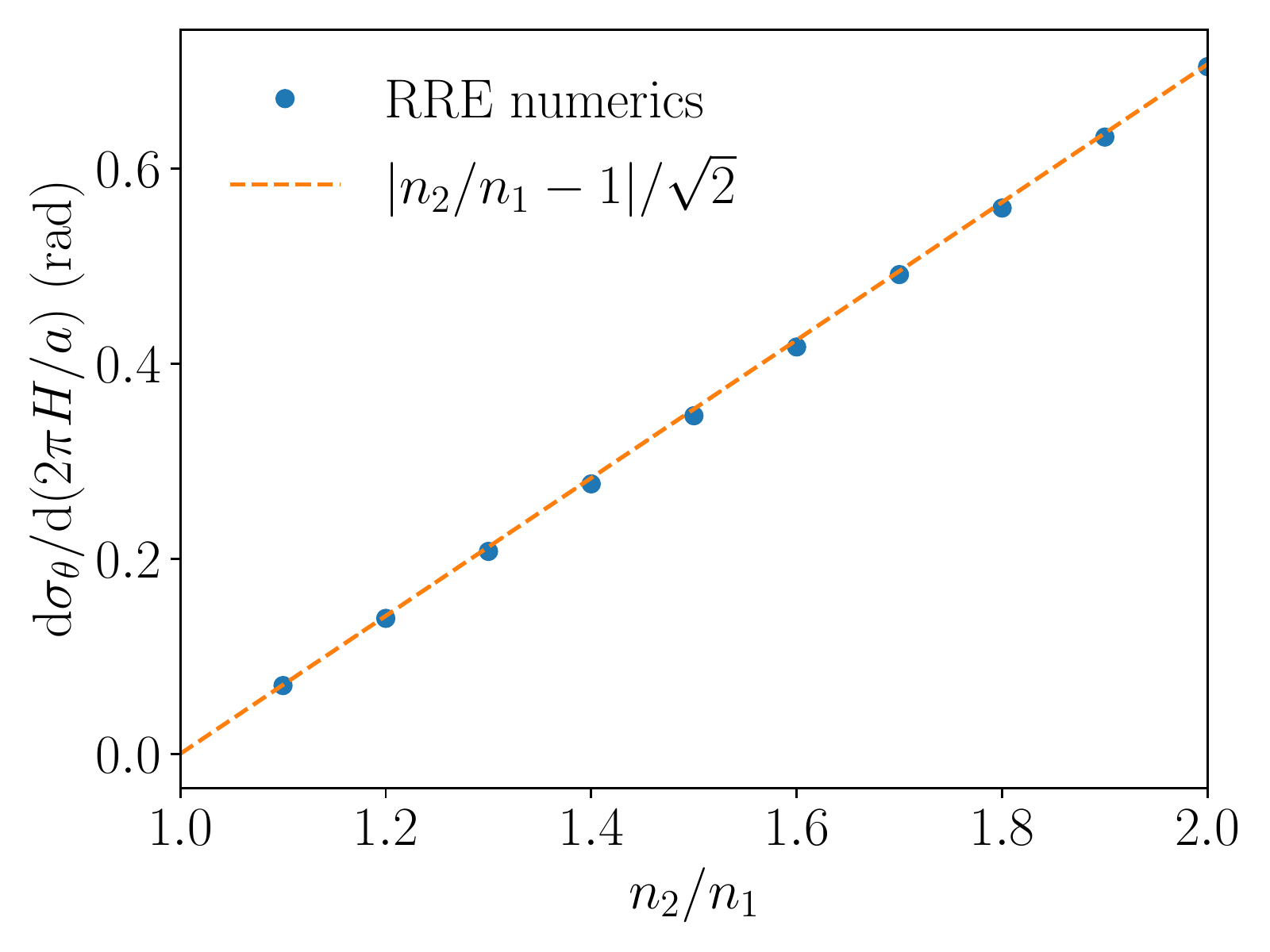}
\caption{Derivative of the angular width of significant diffraction with respect to $2 \pi H / a$ as a function of $n_2 / n_1$. Blue circles: simulation data obtained by fitting a linear function to $\sigma_\theta$ obtained with the reduced Rayleigh equation data for sinusoidal profiles from Fig.~\ref{fig:4} (i.e. fits to the dashed lines from Fig.~\ref{fig:4}). Dashed orange line: analytical expression found in Eq.~(\ref{eq:sigma:scaling:sin}) for sinusoidal profiles (not a fit).}
\label{fig:dsigma}
\end{figure}
%-------------------------------------------------------------------------
Figure~ \ref{fig:4} shows the behavior of $\sigma_\theta$ is indeed linear with $2 \pi H /a$ over a relatively wide range of validity. In addition, Figure~\ref{fig:dsigma} shows the excellent agreement for the derivative of $\sigma_\theta$ with respect to $2 \pi H / a$ between the numerical results obtained by solving the reduced Rayleigh equations and the expression, Eq.~(\ref{eq:sigma:scaling:sin}), derived analytically without any adjustable parameter.

\subsection{Remarks} 
We would like to make two remarks relative to the derivation in this appendix. First, we have made the assumption that the surface profile is small compared to the wavelength to simplify the expression of $\Cie{K}{12}{--}$. The transmission amplitude resulting from this assumption yields the so-called small amplitude perturbation theory to first order. It is a combination of the Born approximation and of the small amplitude assumption. The grating profile treated in this paper \emph{do not} satisfy the small amplitude assumption. Indeed, the amplitude of the most relaxed profile is of the order of a few wavelengths. Nevertheless, the result obtained for the width of significant scattering as a function of $2\pi H / a$ seems to be robust for profile of amplitude of the order of several wavelengths as shown in Fig.~\ref{fig:4} numerically for sinusoidal profiles and experimentally. This suggests that the small amplitude assumption may not be necessary to derive this scaling law. However, we have preferred here to content ourselves with a simple derivation rather than to enter into a more advanced analysis. The second remark concerns the applicability of the small amplitude derivation of $\sigma_\theta$ to sinusoidal profiles. The attentive reader has probably noticed that we have not used the analytical result Eq.~(\ref{eq:Born:sinus}) in the derivation of Appendix~\ref{appendix:angular_window}. The reason is two-fold. First, the derivation stays more general as no assumption is made on the specific shape of the profile. Second, the small amplitude assumption together with the sinusoidal profile actually leads to a rather limited range of validity. Indeed, the result would be significant for $\sigma_\theta < \theta^{(1)}$. The reason for this being that for a sinusoidal profile we have $\Cie{F}{\zeta}{(\ell)} \propto \delta_{|\ell|,1}$, in other words all the diffraction efficiencies vanish except for $\ell = 0, 1$ and $-1$. The fact that the scaling seems to be valid well beyond $\sigma_\theta < \theta^{(1)}$ numerically is quite remarkable and suggests, again, that an alternative derivation, based on the Born approximation but without the small amplitude assumption should be possible.

% --------------------------------------------------------------------
% BIBLIOGRAPHY
% --------------------------------------------------------------------
%
% \section*{References}
% Using BibTeX:
\bibliography{paper2021-11}

%merlin.mbs apsrev4-1.bst 2010-07-25 4.21a (PWD, AO, DPC) hacked
%Control: key (0)
%Control: author (8) initials jnrlst
%Control: editor formatted (1) identically to author
%Control: production of article title (-1) disabled
%Control: page (0) single
%Control: year (1) truncated
%Control: production of eprint (0) enabled
\begin{thebibliography}{19}%
\makeatletter
\providecommand \@ifxundefined [1]{%
 \@ifx{#1\undefined}
}%
\providecommand \@ifnum [1]{%
 \ifnum #1\expandafter \@firstoftwo
 \else \expandafter \@secondoftwo
 \fi
}%
\providecommand \@ifx [1]{%
 \ifx #1\expandafter \@firstoftwo
 \else \expandafter \@secondoftwo
 \fi
}%
\providecommand \natexlab [1]{#1}%
\providecommand \enquote  [1]{``#1''}%
\providecommand \bibnamefont  [1]{#1}%
\providecommand \bibfnamefont [1]{#1}%
\providecommand \citenamefont [1]{#1}%
\providecommand \href@noop [0]{\@secondoftwo}%
\providecommand \href [0]{\begingroup \@sanitize@url \@href}%
\providecommand \@href[1]{\@@startlink{#1}\@@href}%
\providecommand \@@href[1]{\endgroup#1\@@endlink}%
\providecommand \@sanitize@url [0]{\catcode `\\12\catcode `\$12\catcode
  `\&12\catcode `\#12\catcode `\^12\catcode `\_12\catcode `\%12\relax}%
\providecommand \@@startlink[1]{}%
\providecommand \@@endlink[0]{}%
\providecommand \url  [0]{\begingroup\@sanitize@url \@url }%
\providecommand \@url [1]{\endgroup\@href {#1}{\urlprefix }}%
\providecommand \urlprefix  [0]{URL }%
\providecommand \Eprint [0]{\href }%
\providecommand \doibase [0]{http://dx.doi.org/}%
\providecommand \selectlanguage [0]{\@gobble}%
\providecommand \bibinfo  [0]{\@secondoftwo}%
\providecommand \bibfield  [0]{\@secondoftwo}%
\providecommand \translation [1]{[#1]}%
\providecommand \BibitemOpen [0]{}%
\providecommand \bibitemStop [0]{}%
\providecommand \bibitemNoStop [0]{.\EOS\space}%
\providecommand \EOS [0]{\spacefactor3000\relax}%
\providecommand \BibitemShut  [1]{\csname bibitem#1\endcsname}%
\let\auto@bib@innerbib\@empty
%</preamble>
\bibitem [{\citenamefont {Su}\ \emph {et~al.}(2018)\citenamefont {Su},
  \citenamefont {Chu}, \citenamefont {Sun},\ and\ \citenamefont
  {Tsai}}]{Su2018}%
  \BibitemOpen
  \bibfield  {author} {\bibinfo {author} {\bibfnamefont {V.-C.}\ \bibnamefont
  {Su}}, \bibinfo {author} {\bibfnamefont {C.~H.}\ \bibnamefont {Chu}},
  \bibinfo {author} {\bibfnamefont {G.}~\bibnamefont {Sun}}, \ and\ \bibinfo
  {author} {\bibfnamefont {D.~P.}\ \bibnamefont {Tsai}},\ }\href@noop {}
  {\bibfield  {journal} {\bibinfo  {journal} {Opt. Express}\ }\textbf {\bibinfo
  {volume} {26}},\ \bibinfo {pages} {13148–13182} (\bibinfo {year}
  {2018})}\BibitemShut {NoStop}%
\bibitem [{\citenamefont {Turbil}\ \emph {et~al.}(2019)\citenamefont {Turbil},
  \citenamefont {Yoo}, \citenamefont {Simonsen}, \citenamefont {Teisseire},
  \citenamefont {Gozhyk},\ and\ \citenamefont {Garcia-Caurel}}]{Turbil2019}%
  \BibitemOpen
  \bibfield  {author} {\bibinfo {author} {\bibfnamefont {C.}~\bibnamefont
  {Turbil}}, \bibinfo {author} {\bibfnamefont {T.~S.~H.}\ \bibnamefont {Yoo}},
  \bibinfo {author} {\bibfnamefont {I.}~\bibnamefont {Simonsen}}, \bibinfo
  {author} {\bibfnamefont {J.}~\bibnamefont {Teisseire}}, \bibinfo {author}
  {\bibfnamefont {I.}~\bibnamefont {Gozhyk}}, \ and\ \bibinfo {author}
  {\bibfnamefont {E.}~\bibnamefont {Garcia-Caurel}},\ }\href@noop {} {\bibfield
   {journal} {\bibinfo  {journal} {Appl. Opt.}\ }\textbf {\bibinfo {volume}
  {58}},\ \bibinfo {pages} {9267} (\bibinfo {year} {2019})}\BibitemShut
  {NoStop}%
\bibitem [{\citenamefont {Chou}\ \emph {et~al.}(1996)\citenamefont {Chou},
  \citenamefont {Krauss},\ and\ \citenamefont {Renstrom}}]{Chou}%
  \BibitemOpen
  \bibfield  {author} {\bibinfo {author} {\bibfnamefont {S.~Y.}\ \bibnamefont
  {Chou}}, \bibinfo {author} {\bibfnamefont {P.~R.}\ \bibnamefont {Krauss}}, \
  and\ \bibinfo {author} {\bibfnamefont {P.~J.}\ \bibnamefont {Renstrom}},\
  }\href@noop {} {\bibfield  {journal} {\bibinfo  {journal} {Journal of Vacuum
  Science \& Technology B: Microelectronics and Nanometer Structures
  Processing, Measurement, and Phenomena}\ } (\bibinfo {year}
  {1996})}\BibitemShut {NoStop}%
\bibitem [{\citenamefont {Dubov}\ \emph {et~al.}(2013)\citenamefont {Dubov},
  \citenamefont {Perez-Toralla}, \citenamefont {Letailleur}, \citenamefont
  {Barthel},\ and\ \citenamefont {Teisseire}}]{Dubov}%
  \BibitemOpen
  \bibfield  {author} {\bibinfo {author} {\bibfnamefont {A.~L.}\ \bibnamefont
  {Dubov}}, \bibinfo {author} {\bibfnamefont {K.}~\bibnamefont
  {Perez-Toralla}}, \bibinfo {author} {\bibfnamefont {A.}~\bibnamefont
  {Letailleur}}, \bibinfo {author} {\bibfnamefont {E.}~\bibnamefont {Barthel}},
  \ and\ \bibinfo {author} {\bibfnamefont {J.}~\bibnamefont {Teisseire}},\
  }\href@noop {} {\bibfield  {journal} {\bibinfo  {journal} {Journal of
  Micromechanics and Microengineering}\ }\textbf {\bibinfo {volume} {23}},\
  \bibinfo {pages} {125013} (\bibinfo {year} {2013})}\BibitemShut {NoStop}%
\bibitem [{\citenamefont {Brudieu}\ \emph {et~al.}(2017)\citenamefont
  {Brudieu}, \citenamefont {Gozhyk}, \citenamefont {Clements}, \citenamefont
  {Mazoyer}, \citenamefont {Gacoin},\ and\ \citenamefont
  {Teisseire}}]{Brudieu2017}%
  \BibitemOpen
  \bibfield  {author} {\bibinfo {author} {\bibfnamefont {B.}~\bibnamefont
  {Brudieu}}, \bibinfo {author} {\bibfnamefont {I.}~\bibnamefont {Gozhyk}},
  \bibinfo {author} {\bibfnamefont {W.~R.}\ \bibnamefont {Clements}}, \bibinfo
  {author} {\bibfnamefont {S.}~\bibnamefont {Mazoyer}}, \bibinfo {author}
  {\bibfnamefont {T.}~\bibnamefont {Gacoin}}, \ and\ \bibinfo {author}
  {\bibfnamefont {J.}~\bibnamefont {Teisseire}},\ }\href {\doibase
  10.1063/1.4992059} {\bibfield  {journal} {\bibinfo  {journal} {AIP Advances}\
  }\textbf {\bibinfo {volume} {7}},\ \bibinfo {pages} {085215} (\bibinfo {year}
  {2017})}\BibitemShut {NoStop}%
\bibitem [{\citenamefont {Teisseire}\ \emph {et~al.}(2011)\citenamefont
  {Teisseire}, \citenamefont {Revaux}, \citenamefont {Foresti},\ and\
  \citenamefont {Barthel}}]{Teisseire2011}%
  \BibitemOpen
  \bibfield  {author} {\bibinfo {author} {\bibfnamefont {J.}~\bibnamefont
  {Teisseire}}, \bibinfo {author} {\bibfnamefont {A.}~\bibnamefont {Revaux}},
  \bibinfo {author} {\bibfnamefont {M.}~\bibnamefont {Foresti}}, \ and\
  \bibinfo {author} {\bibfnamefont {E.}~\bibnamefont {Barthel}},\ }\href
  {\doibase 10.1063/1.3535614} {\bibfield  {journal} {\bibinfo  {journal}
  {Applied Physics Letters}\ }\textbf {\bibinfo {volume} {98}},\ \bibinfo
  {pages} {013106} (\bibinfo {year} {2011})}\BibitemShut {NoStop}%
\bibitem [{\citenamefont {Kretschmann}\ and\ \citenamefont
  {Maradudin}(2002)}]{kretschmann2002}%
  \BibitemOpen
  \bibfield  {author} {\bibinfo {author} {\bibfnamefont {M.}~\bibnamefont
  {Kretschmann}}\ and\ \bibinfo {author} {\bibfnamefont {A.~A.}\ \bibnamefont
  {Maradudin}},\ }\href@noop {} {\bibfield  {journal} {\bibinfo  {journal}
  {Phys. Rev. B}\ }\textbf {\bibinfo {volume} {66}},\ \bibinfo {pages} {245408}
  (\bibinfo {year} {2002})}\BibitemShut {NoStop}%
\bibitem [{\citenamefont {Banon}(2018)}]{banon:thesis}%
  \BibitemOpen
  \bibfield  {author} {\bibinfo {author} {\bibfnamefont {J.-P.}\ \bibnamefont
  {Banon}},\ }\emph {\bibinfo {title} {On the simulation of electromagnetic
  wave scattering by periodic and randomly rough layered structures based on
  the reduced {R}ayleigh equations : theory, numerical analysis and
  applications}},\ \href@noop {} {Ph.D. thesis},\ \bibinfo  {school} {Norwegian
  University of Science and Technology}, \bibinfo {address} {Trondheim, Norway}
  (\bibinfo {year} {2018})\BibitemShut {NoStop}%
\bibitem [{\citenamefont {Toigo}\ \emph {et~al.}(1977)\citenamefont {Toigo},
  \citenamefont {Marvin}, \citenamefont {Celli},\ and\ \citenamefont
  {Hill}}]{Toigo:77}%
  \BibitemOpen
  \bibfield  {author} {\bibinfo {author} {\bibfnamefont {F.}~\bibnamefont
  {Toigo}}, \bibinfo {author} {\bibfnamefont {A.}~\bibnamefont {Marvin}},
  \bibinfo {author} {\bibfnamefont {V.}~\bibnamefont {Celli}}, \ and\ \bibinfo
  {author} {\bibfnamefont {N.~R.}\ \bibnamefont {Hill}},\ }\href@noop {}
  {\bibfield  {journal} {\bibinfo  {journal} {Phys. Rev. B}\ }\textbf {\bibinfo
  {volume} {15}},\ \bibinfo {pages} {5618} (\bibinfo {year}
  {1977})}\BibitemShut {NoStop}%
\bibitem [{\citenamefont {Soubret}\ \emph {et~al.}(2001)\citenamefont
  {Soubret}, \citenamefont {Berginc},\ and\ \citenamefont
  {Bourrely}}]{soubret1}%
  \BibitemOpen
  \bibfield  {author} {\bibinfo {author} {\bibfnamefont {A.}~\bibnamefont
  {Soubret}}, \bibinfo {author} {\bibfnamefont {G.}~\bibnamefont {Berginc}}, \
  and\ \bibinfo {author} {\bibfnamefont {C.}~\bibnamefont {Bourrely}},\
  }\href@noop {} {\bibfield  {journal} {\bibinfo  {journal} {Phys. Rev. B}\
  }\textbf {\bibinfo {volume} {63}},\ \bibinfo {pages} {245411} (\bibinfo
  {year} {2001})}\BibitemShut {NoStop}%
\bibitem [{\citenamefont {Hetland}\ \emph {et~al.}(2017)\citenamefont
  {Hetland}, \citenamefont {Maradudin}, \citenamefont {Nordam}, \citenamefont
  {Letnes},\ and\ \citenamefont {Simonsen}}]{hetland:17}%
  \BibitemOpen
  \bibfield  {author} {\bibinfo {author} {\bibfnamefont {{\O}.~S.}\
  \bibnamefont {Hetland}}, \bibinfo {author} {\bibfnamefont {A.~A.}\
  \bibnamefont {Maradudin}}, \bibinfo {author} {\bibfnamefont {T.}~\bibnamefont
  {Nordam}}, \bibinfo {author} {\bibfnamefont {P.~A.}\ \bibnamefont {Letnes}},
  \ and\ \bibinfo {author} {\bibfnamefont {I.}~\bibnamefont {Simonsen}},\
  }\href@noop {} {\bibfield  {journal} {\bibinfo  {journal} {Phys. Rev. A}\
  }\textbf {\bibinfo {volume} {95}},\ \bibinfo {pages} {043808} (\bibinfo
  {year} {2017})}\BibitemShut {NoStop}%
\bibitem [{\citenamefont {Banon}\ \emph {et~al.}(2019)\citenamefont {Banon},
  \citenamefont {Hetland},\ and\ \citenamefont {Simonsen}}]{Banon:2018}%
  \BibitemOpen
  \bibfield  {author} {\bibinfo {author} {\bibfnamefont {J.-P.}\ \bibnamefont
  {Banon}}, \bibinfo {author} {\bibfnamefont {{\O}.~S.}\ \bibnamefont
  {Hetland}}, \ and\ \bibinfo {author} {\bibfnamefont {I.}~\bibnamefont
  {Simonsen}},\ }\href {\doibase 10.1103/PhysRevA.99.023834} {\bibfield
  {journal} {\bibinfo  {journal} {Phys. Rev. A}\ }\textbf {\bibinfo {volume}
  {99}},\ \bibinfo {pages} {023834} (\bibinfo {year} {2019})}\BibitemShut
  {NoStop}%
\bibitem [{\citenamefont {Gonz\'{a}lez-Alcalde}\ \emph
  {et~al.}(2016)\citenamefont {Gonz\'{a}lez-Alcalde}, \citenamefont {Banon},
  \citenamefont {Hetland}, \citenamefont {Maradudin}, \citenamefont
  {M\'{e}ndez}, \citenamefont {Nordam},\ and\ \citenamefont
  {Simonsen}}]{Gonzalez-Alcalde:16}%
  \BibitemOpen
  \bibfield  {author} {\bibinfo {author} {\bibfnamefont {A.~K.}\ \bibnamefont
  {Gonz\'{a}lez-Alcalde}}, \bibinfo {author} {\bibfnamefont {J.-P.}\
  \bibnamefont {Banon}}, \bibinfo {author} {\bibfnamefont {{\O}.~S.}\
  \bibnamefont {Hetland}}, \bibinfo {author} {\bibfnamefont {A.~A.}\
  \bibnamefont {Maradudin}}, \bibinfo {author} {\bibfnamefont {E.~R.}\
  \bibnamefont {M\'{e}ndez}}, \bibinfo {author} {\bibfnamefont
  {T.}~\bibnamefont {Nordam}}, \ and\ \bibinfo {author} {\bibfnamefont
  {I.}~\bibnamefont {Simonsen}},\ }\href@noop {} {\bibfield  {journal}
  {\bibinfo  {journal} {Opt. Express}\ }\textbf {\bibinfo {volume} {24}},\
  \bibinfo {pages} {25995} (\bibinfo {year} {2016})}\BibitemShut {NoStop}%
\bibitem [{\citenamefont {Banon}\ \emph {et~al.}(2018)\citenamefont {Banon},
  \citenamefont {Hetland},\ and\ \citenamefont {Simonsen}}]{Banon:17:2}%
  \BibitemOpen
  \bibfield  {author} {\bibinfo {author} {\bibfnamefont {J.-P.}\ \bibnamefont
  {Banon}}, \bibinfo {author} {\bibfnamefont {{\O}.~S.}\ \bibnamefont
  {Hetland}}, \ and\ \bibinfo {author} {\bibfnamefont {I.}~\bibnamefont
  {Simonsen}},\ }\href@noop {} {\bibfield  {journal} {\bibinfo  {journal} {Ann.
  Phys.}\ }\textbf {\bibinfo {volume} {389}},\ \bibinfo {pages} {352} (\bibinfo
  {year} {2018})}\BibitemShut {NoStop}%
\bibitem [{\citenamefont {Simonsen}\ \emph {et~al.}(2009)\citenamefont
  {Simonsen}, \citenamefont {Larsen}, \citenamefont {Andreassen}, \citenamefont
  {Ommundsen},\ and\ \citenamefont {Nord-Varhaug}}]{Simonsen2002-1}%
  \BibitemOpen
  \bibfield  {author} {\bibinfo {author} {\bibfnamefont {I.}~\bibnamefont
  {Simonsen}}, \bibinfo {author} {\bibfnamefont {A.}~\bibnamefont {Larsen}},
  \bibinfo {author} {\bibfnamefont {E.}~\bibnamefont {Andreassen}}, \bibinfo
  {author} {\bibfnamefont {E.}~\bibnamefont {Ommundsen}}, \ and\ \bibinfo
  {author} {\bibfnamefont {K.}~\bibnamefont {Nord-Varhaug}},\ }\href {\doibase
  10.1103/PhysRevA.79.063813} {\bibfield  {journal} {\bibinfo  {journal} {Phys.
  Rev. A}\ }\textbf {\bibinfo {volume} {79}},\ \bibinfo {pages} {063813}
  (\bibinfo {year} {2009})}\BibitemShut {NoStop}%
\bibitem [{\citenamefont {Gozhyk}\ \emph {et~al.}(2023)\citenamefont {Gozhyk},
  \citenamefont {Turbil}, \citenamefont {Garcia},\ and\ \citenamefont
  {Obein}}]{Colette_NCS}%
  \BibitemOpen
  \bibfield  {author} {\bibinfo {author} {\bibfnamefont {I.}~\bibnamefont
  {Gozhyk}}, \bibinfo {author} {\bibfnamefont {C.}~\bibnamefont {Turbil}},
  \bibinfo {author} {\bibfnamefont {E.}~\bibnamefont {Garcia}}, \ and\ \bibinfo
  {author} {\bibfnamefont {G.}~\bibnamefont {Obein}},\ }\href@noop {}
  {\bibfield  {journal} {\bibinfo  {journal} {to be submitted to Journal of
  Optics}\ } (\bibinfo {year} {2023})}\BibitemShut {NoStop}%
\bibitem [{\citenamefont {Stover}(2015)}]{Stover}%
  \BibitemOpen
  \bibfield  {author} {\bibinfo {author} {\bibfnamefont {J.}~\bibnamefont
  {Stover}},\ }\href@noop {} {\emph {\bibinfo {title} {Optical scattering
  measurement and analysis,}}}\ (\bibinfo {year} {2015})\BibitemShut {NoStop}%
\bibitem [{\citenamefont {Petit}(1980)}]{Petit:1980}%
  \BibitemOpen
  \bibinfo {editor} {\bibfnamefont {R.}~\bibnamefont {Petit}},\ ed.,\ \href
  {\doibase 10.1007/978-3-642-81500-3} {\emph {\bibinfo {title}
  {Electromagnetic Theory of Gratings}}}\ (\bibinfo  {publisher} {Springer
  Berlin Heidelberg},\ \bibinfo {year} {1980})\BibitemShut {NoStop}%
\bibitem [{\citenamefont {Maradudin}(1983)}]{Maradudin:83}%
  \BibitemOpen
  \bibfield  {author} {\bibinfo {author} {\bibfnamefont {A.~A.}\ \bibnamefont
  {Maradudin}},\ }\href@noop {} {\bibfield  {journal} {\bibinfo  {journal} {J.
  Opt. Soc. Am.}\ }\textbf {\bibinfo {volume} {73}},\ \bibinfo {pages} {759}
  (\bibinfo {year} {1983})}\BibitemShut {NoStop}%
\end{thebibliography}%

%----------------------------------------------------------------------
\end{document}